\documentclass[prd,superscriptaddress,nofootinbib,twocolumn,showpacs,floatfix]{revtex4}

\usepackage{graphics,graphicx}
\usepackage{amsmath,amssymb}
\usepackage{natbib}
\usepackage{bm}
\usepackage{url}
\usepackage{color}
\usepackage{dcolumn}
\usepackage{dashundergaps}
\usepackage{dashrule}

\def\be{\begin{equation}}
\def\ee{\end{equation}}
\def\ba{\begin{eqnarray}}
\def\ea{\end{eqnarray}}
\def\nn{\nonumber}

\def\bc{\begin{center}}
\def\ec{\end{center}}

\def\fnl{f_{\rm NL}}
\def\l{\ell}

\begin{document}

\title{Application of cross correlations between CMB and large scale structure
  to constraints on the primordial non-Gaussianity}
 
\author{Yoshitaka Takeuchi}
\email{yositaka@a.phys.nagoya-u.ac.jp}
\affiliation{Department of Physics, Nagoya University, Chikusa-ku, Nagoya,  464-8602, Japan}
\author{Kiyotomo Ichiki}
\email{ichiki@a.phys.nagoya-u.ac.jp}
\affiliation{Department of Physics, Nagoya University, Chikusa-ku, Nagoya,  464-8602, Japan}
\author{Takahiko Matsubara}
\email{taka@a.phys.nagoya-u.ac.jp}
\affiliation{Kobayashi-Maskawa Institute for the Origin of Particles and the Universe, Nagoya
University, Chikusa-ku, Nagoya,  464-8602, Japan}

\date{\today}

\pacs{98.80.-k, 98.62.Sb, 98.65.-r}
         
\begin{abstract}
The primordial non-Gaussianity of local type affects the clustering of dark
matter halos, and the planned deep and wide photometric surveys are suitable
for examining this class of non-Gaussianity.  In our previous paper, we
investigated the constraint from the cross correlation between CMB lensing
potential and galaxy angular distribution on the primordial non-Gaussianity,
without taking into account redshift slicing.  To improve our previous
analysis, in this paper, we add the galaxy lensing shear into our analysis and
take into account redshift slicing to follow the redshift evolution of the
clustering.  By calculating 81 power spectra and using the Fisher matrix
method, we find that the constraint on the primordial non-Gaussianity can be
improved from $\Delta f_{\rm NL} \sim 5.4$ to $5.1$ by including the
galaxy-galaxy lensing shear cross correlations expected from the Hyper
Suprime-Cam survey (HSC), in comparison with the constraint without any cross
correlations.  Moreover, the constraint can go down to $\Delta f_{\rm NL} \sim
4.8$ by including the galaxy-CMB lensing cross correlations from the ACTPol
and Planck experiments.
\end{abstract}

\maketitle
 

\section{Introduction} \label{sec:intro}

Primordial non-Gaussianities have been intensively discussed, because any
detection of them offers an important window into the early Universe. Standard
single-field slow-roll inflation models predict small non-Gaussianity
\cite{Mukhanov81,Allen87,Gangui94,Maldacena03,Acquaviva03}, so that the
detection of large non-Gaussianity may turn down the standard model and
suggest physics beyond the standard model. The most popular method to hunt for
primordial non-Gaussianities is to measure higher order correlation functions
of the cosmic microwave background (CMB) anisotropies such as the the
three-point correlation function (bispectrum) and the four-point one
(trispectrum). Non-vanishing signals of these higher order correlations may
predict the presence of primordial non-Gaussianities.
\cite{Komatsu03,Spergel07,Creminelli07,Komatsu09,Smith09,Komatsu11}.

Recently, many studies have revealed that the primordial non-Gaussianity of
local type affects the large-scale structure (LSS) through the clustering of
dark matter halos, and shown the modification of the halo mass function
\cite{LoVerde08,Matarrese00,Verde01,DAmico11,Grossi09,Yokoyama11a} and the
halo bias
\cite{Slosar08,Dalal08,Afshordi08,Matarrese08,Yokoyama11b,Gong11,Desjacques09},
both by numerical simulations and by analytic calculations.  Among several
types of primordial non-Gaussianities, the local-type non-Gaussianity induces
strong scale-dependence of the halo bias
\cite{Slosar08,Dalal08,Afshordi08,Matarrese08}. This scale-dependent bias is
also powerful tool to constrain the primordial non-Gaussianity from the
observations of LSS independently of the method with CMB higher order
correlation functions.

Using this scale-dependent feature in the halo bias, some measurements of the
primordial non-Gaussianity have been done in Refs.
\cite{Slosar08,Afshordi08,Xia11}. It is expected that future wide and deep
surveys, such as Subaru Hyper Suprime-Cam (HSC) survey \cite{HSC}, Dark Energy
Survey (DES) \cite{DES}, Large Synoptic Survey Telescope (LSST) \cite{LSST},
and so on, will put tighter constraints on the primordial non-Gaussianity,
which are expected to be comparable to those from the future CMB observations
or more (e.g., \cite{Carbone08,Carbone10,Oguri11,Cunha10}).

To get a tighter constraint on the primordial non-Gaussianity, it is important
to distinguish the signature of the primordial non-Gaussianity from the other
effects more precisely, especially from the linear bias in the limited survey
area. Thus, it is better for this purpose to combine unbiased observables
which are sensitive to the matter power spectrum in the near Universe, such as
weak gravitational lensing of CMB and galaxy lensing shear.

Cross correlations with complementary probes are expected to provide
additional information on top of their respective auto correlations.  The
cross correlation between CMB and LSS is known to give information about the
integrated Sachs-Wolfe (ISW) effect of CMB temperature anisotropy, which
generates the secondary anisotropies due to the time variation of the
gravitational potential \cite{Sachs67}.  Moreover, CMB lensing also has a
correlation with LSS since the gravitational lensing of CMB is induced by the
gravitational potential produced by LSS, and the galaxy lensing shear as well
\cite{Jeong09}.

Since the effects of the CMB lensing are imprinted on small scales, the
Wilkinson Microwave Anisotropy Probe (WMAP) satellite has not detected the CMB
lensing directly. Recent ground based CMB experiments such as Atacama Cosmology
Telescope (ACT), announced the detection of the CMB lensing signal, and an
ongoing CMB observation by Planck \cite{Planck} or various ground based
experiments are expected to detect this signal more precisely. Therefore, we
can be sure that the observation of CMB lensing will increase its accuracy and
play an important role in observational cosmology in near future surveys.

In this paper we consider prospects for constraining the primordial
non-Gaussianity through the scale-dependent bias from the near future surveys
taking into account all auto and cross correlations among CMB, galaxy
distribution and galaxy lensing shear.  We shall show that galaxy-CMB lensing
and galaxy-galaxy lensing cross correlations are particularly fruitful and
allow us to improve the constraints on the primordial non-Gaussianity.

This paper is organized as follows.  In section \ref{sec:NG} we briefly review
the effects of the primordial non-Gaussianity for large-scale structure and
show the modifications to the mass function and bias of dark matter halos.  In
section \ref{sec:cls} we summarize the observables employed in our analysis
and describe the angular power spectra of the cross correlations between CMB
and LSS. In section \ref{sec:survey} we mention the survey design and
photometric redshift systematics employed in our analysis.  In section
\ref{sec:forecasts} we explain the method and the setup of our analysis.  In
section \ref{sec:result} we show the results particularly focusing on the
constraints on the primordial non-Gaussianity.  In section
\ref{sec:discussion} we discuss the effects of the photometric redshift
systematics and the massive neutrino on the constraints.  Then we assess the
contribution of the cross correlations for the constraints. Moreover, we
compare the constraints from some different surveys. Finally, in section
\ref{sec:summary} we summarize our conclusions. Through this paper we assume a
spatially flat Universe for simplicity.


\section{Primordial non-Gaussianity} \label{sec:NG}

Deviations from Gaussian initial conditions are commonly parameterized in
terms of the dimensionless parameter $\fnl$ and the primordial non-Gaussianity
of the local form is described as \cite{Komatsu01,Verde00,Gangui94}
\begin{equation}
  \Phi = \phi + f_{\rm NL}(\phi^2 - \langle \phi^2 \rangle ) ,
\label{eq:NG_local}
\end{equation}
where $\Phi$ is the curvature perturbation and $\phi$ is a Gaussian random
field. On the subhorizon scale, $\Phi$ is related to the Newtonian
gravitational potential $\Psi$ as $\Phi = - \Psi$.

The existence of the primordial non-Gaussianity, $\fnl \neq 0$, indicates that
the initial density field is positively or negatively skewed.  Furthermore,
the fact that the non-Gaussianity affects the clustering of dark matter halos
or galaxies allows us to constrain on the non-Gaussianity from large-scale
structure surveys.  In particular, the local-type non-Gaussianity described by
Eq.~(\ref{eq:NG_local}) induces a scale-dependent enhancement of the
halo/galaxy power spectrum.

In the presence of the primordial non-Gaussianity, the mass function of
clustering halos is modified and we adopt a non-Gaussian correction factor of
the halo mass function based on the Edgeworth expansion \cite{LoVerde08}:
\begin{equation}
  \frac{dn/dM}{dn_{\rm G}/dM} =
  1 + \frac{\sigma_M S_3}{6}(\nu^2 - 3\nu) 
  -\frac{1}{6}\frac{d(\sigma_M S_3)}{d \ln \nu} 
  \left( \nu - \frac{1}{\nu} \right) ,
\label{eq:NG_MF}
\end{equation}
where $S_3$ is the skewness of the density field which is proportional to
$\fnl$, $\sigma_M=\sigma_M(M,z)$ is the rms of the linear density field
smoothed on mass scale $M$, $\nu$ is defined as $\nu = \delta_{\rm
  c}/\sigma_M$ and $\delta_{\rm c} \sim 1.68$ is the critical linear
overdensity. $dn/dM$ is the mass function in the non-Gaussian case and
$dn_{\rm G}/dM$ is the one in the Gaussian case. For the Gaussian one, we
adopt a model of Warren $et~al$ \cite{Warren06}.

Recent studies have shown that the local-type primordial non-Gaussianity
produces a scale-dependent enhancement of the clustering of halo on
large-scales,
\begin{equation}
  P_{\rm g}(M,z,k) = b_{\rm G}^2(M,z)P(k,z) 
  \rightarrow [b_{\rm G} + \Delta b(M,z,k)]^2P(k,z) ,
\label{eq:NG_Bias}
\end{equation}
where $P_{\rm g}(M,z,k)$ is the galaxy power spectrum, $P(k,z)$ is the matter
power spectrum, $b_{\rm G}(M,z)$ is the bias in the Gaussian case and $\Delta
b(M,z,k)$ is the non-Gaussian correction of the halo bias described as
\cite{Desjacques09}
\begin{multline}
  \Delta b(M,z,k) = 
  \frac{3 \Omega_{\rm m, 0}H_0^2}{k^2 T(k) D(z)}
  f_{\rm NL}\delta_{\rm c}(b_{\rm G}(M,z)-1) \\
  - \frac{\nu}{\delta_{\rm c}}\frac{d}{d\nu}
  \left(\frac{dn/dM}{dn_{\rm G}/dM}\right) .
\label{eq:deltabk}
\end{multline}
Here, $\Omega_{\rm m, 0}$ and $H_0$ are the matter energy density normalized
by the critical density and the Hubble parameter at present, $D(z)$ is the
linear growth rate normalized to the scale factor $a$ in the matter-dominant
era and $T(k)$ is the transfer function of linear matter density
fluctuations. For the halo bias in the Gaussian case $b_{\rm G}(M,z)$, we
assumed the form presented by Sheth $et~al$ \cite{Sheth01}.


\section{Angular power spectrum} \label{sec:cls}

Here we briefly review the auto and cross correlations of various cosmological
observables and their angular power spectra. In this paper we take into
account CMB temperature ($T$) anisotropies, $E$-mode polarization ($E$) and
CMB lensing potential ($\psi$) for CMB observables and galaxy distribution (g)
and weak lensing shear ($\gamma$) for LSS observables.


\subsection{Integrated Sachs-Wolfe (ISW) effect}

The decay of the Newtonian potential due to the presence of dark energy
produces a differential gravitational redshift, and this effect is called the
late-time integrated Sachs-Wolfe (ISW) effect. In a flat Universe the presence
of ISW effect is a signature of dark energy, and induces a non-vanishing cross
correlation between CMB temperature and large-scale structure measurements,
such as galaxy distribution, weak lensing field and so on.

The contribution of the ISW effect to the CMB temperature field
$\Theta(\hat{\bm n})$ can be written as
\begin{equation}
  \Theta_{\rm ISW}(\hat{\bm n}) 
  = \frac{\Delta T_{\rm ISW}(\hat{\bm n})}{T_{\rm CMB}}
  = - \frac{2}{T_{\rm CMB}} \int_0^{\chi_{*}} d\chi 
  \dot{\Phi}(\chi \hat{\bm n}, \chi) ,
\end{equation}
where $T_{\rm CMB}$ is the mean temperature of the CMB, $\hat{\bm n}$ is the
direction to the line of sight, $\chi$ is the comoving distance and $\chi_{*}
$ denotes the distance to the last scattering surface.  $\Phi$ is the
gravitational potential and a dot denotes a derivative with respect to the
conformal time.

The angular power spectrum of the cross correlation between CMB temperature
through ISW effect and the other measurements $X$ can be written as
\begin{equation}
  C_\l^{TX} = \frac{2}{\pi} \int k^2 dk P_{\Phi}(k) 
  \Delta_\l^{\rm ISW}(k) \Delta_\l^X(k) ,
\end{equation}
where
\begin{equation}
  \Delta_\l^{\rm ISW}(k) = 
  3\Omega_{\rm m,0} H_0^2  \int_0^{z_{*}} dz \frac{d}{dz}
  \left\{ \frac{D(z)}{a(z)} \right\} T(k) j_\l(k\chi(z)) ,
\end{equation}
$P_{\Phi}(k) \propto k^{n^{\rm s}-4}$ is the primordial power spectrum of
$\Phi$ as a function of the wave number $k$ and $n^{\rm s}$ is the tilt of the
primordial power spectrum. The functions $T(k)$ and $D(z)$ are the transfer
function and the growth rate for linear matter density fluctuations,
respectively, and $j_\l(k\chi)$ is a spherical Bessel function. The kernel
$\Delta_\l^X(k)$ is for the other measurements, namely, CMB lensing potential,
galaxy distribution and weak lensing shear in this paper; $X=\psi, {\rm g},
\gamma$, respectively.

The noise spectra of CMB include detector noise and residual foreground
contamination. Here, we assume the ideal condition that foreground
contamination can be completely removable and include only Gaussian random
detector noise of the form \cite{Knox95}
\begin{equation}
  N_\l^{T,P} = \left[ 
    \sum_\nu \left\{ (\theta_{\rm FWHM}\Delta_{T,P})^{-2} 
    e^{-\l(\l+1)\theta_{\rm FWHM}/8\ln 2} \right\}
    \right]^{-1} ,
\label{eq:noise_cmb}
\end{equation}
where $\theta_{\rm FWHM}$ is the spatial resolution of the beam and
$\Delta_{T,P}$ represents the sensitivity to the temperature and polarization
per pixel, respectively. These values are given for each of the frequency
bands of channels $\nu$ and we show the values for two CMB experiments such as
Planck \cite{Planck} and ACT with new polarization sensitive receiver (ACTPol)
\cite{ACTPol} in Table~\ref{tb:CMB_exp.}.


\subsection{CMB lensing potential}

CMB photons are deflected due to the gravitational potential produced by the
large-scale structure on the way propagating to us (e.g., Ref.~\cite{Lewis06}). The
relationship between the lensed temperature anisotropy $\tilde{T}(\hat{\bm
  n})$ and the unlensed one $T(\hat{\bm n})$ is given by the deflection angle
${\bm d}(\hat{\bm n})$ as $\tilde{T}(\hat{\bm n}) = T(\hat{\bm n}+{\bm d})$.
The deflection is related to the line of sight projection of the gravitational
potential $\Psi (\hat{\bm n},z)$ as ${\bm d}(\hat{\bm n})=\nabla \psi
(\hat{\bm n})$, where
\begin{equation}
  \psi(\hat{\bm n}) = -2 \int_0^{\chi_{*}} d\chi
  \frac{\chi_{*} - \chi}{\chi_{*}\chi} \Psi(\chi\hat{\bm n},\chi) .
\end{equation}
Here, $\psi(\hat{\bm n})$ is the (effective) lensing potential.

The angular power spectrum of the lensing potential can be written as 
\begin{equation}
  C_\l^\psi = \frac{2}{\pi} \int k^2 dk P_{\Phi}(k) [\Delta_\l^\psi (k)]^2 ,
\end{equation}
where
\begin{equation}
  \Delta_\l^\psi(k) = 3\Omega_{\rm m, 0}H_0^2 \int_0^{\chi_{*}} d\chi
  \frac{\chi_{*} - \chi}{\chi_{*}\chi} 
  \frac{T(k)D(z(\chi))}{a(\chi)} j_{\l}(k\chi) .
\end{equation}

The lensing potential can be reconstructed with quadratic statistics in the
temperature and polarization data that are optimized to extract the lensing
signal. To reconstruct the lensing potential, one needs to use the
non-Gaussian nature imprinted into the lensed CMB statistics, and the noise of
the lensing potential can be estimated as the reconstruction error
\cite{Hu02,Okamoto03}. In this paper, we estimate the noise spectrum of
lensing potential $N_\l^\psi$ following the technique developed in
\cite{Okamoto03} optimally combining the temperature and polarization fields.


\begin{table}[t]
\begin{ruledtabular}
\begin{tabular}{ccccccc}
 experiment & 
 $f_{\rm sky}$ & $\nu$ & $\theta_{\rm FWHM}$ & $\Delta_T$ & $\Delta_P$  \\
 &  & [GHz] & [arcmin] & [$\mu$K/pixel] & [$\mu$K/pixel]  \\
\hline
Planck & 0.65 
  & 100 & 9.5' & 6.8 & 10.9   \\
  && 143 & 7.1' & 6.0 & 11.4  \\
  && 217 & 5.0' & 13.1 & 26.7 \\
\hline
ACTPol  & 0.05 
  & 148 & 1.4' & 3.6 & 5.0
\end{tabular}
\end{ruledtabular}
\caption{The specifications for CMB experiments. $f_{\rm sky}$ is the sky
  coverage, $\theta_{\rm FWHM}$ is the beam width at FWHM, $\Delta_T$ and
  $\Delta_P$ represent the sensitivity of each channel to the temperature and
  polarization, respectively, and $\nu$ means the channel frequency. }
\label{tb:CMB_exp.}
\end{table}


\subsection{Galaxy distribution}

\begin{figure}[t]
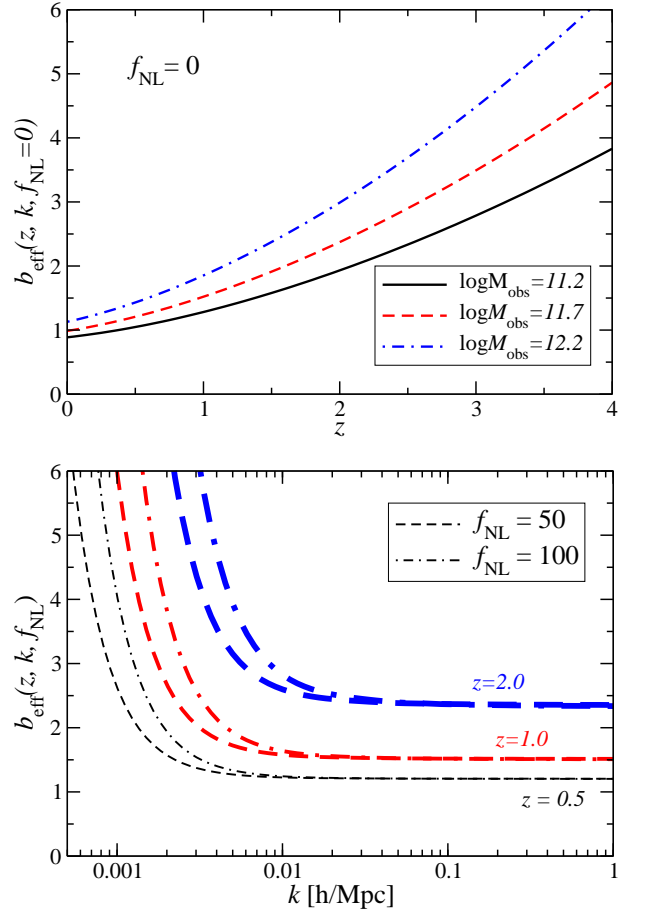

  \bc
  \includegraphics[clip,keepaspectratio=true,width=0.45
    \textwidth]{fig1_beff-z.eps}

  \vspace{8pt}
  \includegraphics[clip,keepaspectratio=true,width=0.45
    \textwidth]{fig1_beff-k.eps}
  \ec
\caption{The effective weighted bias defined in Eq.~(\ref{eq:beff}) as a
  function of redshift $z$ (Top) and as a function of wave number $k$
  (Bottom).  (Top) We plot the Gaussian case ($f_{\rm NL}=0$) for the values
  $\log M_{\rm obs}=11.2,11.7,12.2$, respectively, and the $k$ dependence does
  not appear in this case. (Bottom) We plot the non-Gaussian case ($f_{\rm
    NL}=50, 100$) at redshift $z=0.5,1.0,2.0$ as indicated, and the value
  $\log M_{\rm obs}$ is fixed to be $\log M_{\rm obs}=11.7$.}
\label{fig:beff}
\end{figure}

The luminous sources such as galaxies must be the most obvious tracers of the
large-scale structure in the linear regime, and the projected density contrast
of the galaxies can be written as
\begin{equation}
  \delta_{{\rm g},i}(\hat{\bm n}) = \int_0^{\infty}dz \,
  b_{\rm eff}(z, k)\frac{n_i(z)}{n_i^{\rm A}} \delta(\chi(z)\hat{\bm n}, z),
\end{equation}
where the subscript $i$ represents the $i$-th redshift
bin. $\delta(\chi\hat{\bm n}, z)$ represents the matter density fluctuation,
and $n_i(z)$ and $n_i^{\rm A}$ are the galaxy redshift distribution and the
total number of galaxies per steradian in the $i$-th redshift bin. The
function $b_{\rm eff}(z,k)$ is the weighted effective halo/galaxy bias defined
as
\begin{equation}
  b_{\rm eff}(z, k) = 
  \left[{\int_{M_{\rm obs}}^{\infty}dM \frac{dn}{dM}}\right]^{-1}
       {\int_{M_{\rm obs}}^{\infty}dM b(M,z,k) \frac{dn}{dM}} ,
\label{eq:beff}
\end{equation}
where $M_{\rm obs}$ is the observable mass threshold, which is the minimum
mass of the galaxy we can observe, and we take the value to be $M_{\rm
  obs}=10^{11.7} [h^{-1} {\rm M_{\odot}}]$. The function $b(M,z,k)$ and
$dn/dM$ are the halo mass function and the halo bias in the case of the
non-Gaussian initial condition defined in Eqs.~(\ref{eq:NG_MF}) and
(\ref{eq:NG_Bias}), respectively.  We show the dependence of $b_{\rm
  eff}(z,k)$ on the redshift and the wave number with different mass
thresholds and $f_{\rm NL}$ in Fig.~\ref{fig:beff}.

The angular power spectrum of the galaxy distribution between $i$-th and
$j$-th redshift bins can be written as
\begin{equation}
  C_\l^{{\rm g}_i {\rm g}_j} = 
  \frac{2}{\pi} \int k^2 dk P_\Phi(k)
  \Delta_\l^{{\rm g}_i}(k) \Delta_\l^{{\rm g}_j}(k) ,
\end{equation}
where
\begin{equation}
  \Delta_\l^{{\rm g}_i}(k) = \int dz \,
  b_{\rm eff}(z, k)\frac{n_i(z)}{n_i^{\rm A}}
  T(k)D(z) j_{\l}(k\chi) .
\end{equation}

To estimate the signal-to-noise ratios and errors in parameter determination
for each survey, we need to describe the noise contribution due to the
finiteness in the number of sources associated with source samples. We can
write the noise spectra from the shot noise as
\begin{equation}
  N_\l^{{\rm g}_i{\rm g}_j} = \delta_{ij}\frac{1}{\bar{n}_i} ,
\label{eq:noise_gal}
\end{equation}
where $\bar{n}_i$ is the mean surface density of sources per steradian in
the $i$-th redshift bin.


\subsection{Weak lensing shear}
The weak lensing shear (equivalently the convergence in the weak lensing
limit) is a weighted integral of the density field of sources, which is
directly related to the source galaxy redshift distribution. The average
convergence of a light ray bundle from sources in the $i$-th redshift bin is
written as
\begin{equation}
  \kappa_i(\hat{\bm n}) = 
  \int_0^{\infty} d\chi W_i(\chi) \delta(\chi\hat{\bm n}, \chi) ,
\end{equation}
where $W_i(\chi)$ is the convergence weight function of kernel defined as
\begin{equation}
  W_i(\chi(z)) = \frac{3\Omega_{{\rm m},0}H_0^2 }{2}\frac{\chi(z)}{a(z)}
  \int_z^{\infty} dz{'}\frac{n_i(z{'})}{n_i^{\rm A}}
  \frac{\chi(z{'})-\chi(z)}{\chi(z{'})} .
\end{equation}
Here $\chi(z)$ is the distance to the lens and $\chi(z{'})$ is the one to the
source.

The angular power spectra of the weak lensing shear between $i$-th and $j$-th
redshift bins can be written as
\begin{equation}
  C_\l^{{\gamma}_i {\gamma}_j} = 
  \frac{2}{\pi} \int k^2 dk P_\Phi(k)
  \Delta_\l^{{\gamma}_i}(k) \Delta_\l^{{\gamma}_j}(k) ,
\end{equation}
where
\begin{equation}
  \Delta_\l^{{\gamma}_i}(k) = \int dz \,
  W_i(z)
  T(k)D(z) j_{\l}(k\chi) .
\end{equation}
 The measurement of the shear from galaxy images has some uncertainties, one
 of which mainly comes from the intrinsic shape ellipticities of galaxies. The
 galaxy shapes can be treated statistically through a shape noise contribution
 and the noise spectra can be written as
\begin{equation}
  N_\l^{{\gamma}_i {\gamma}_j} = 
  \delta_{ij}\frac{\sigma_\gamma^2 }{\bar{n}_i} ,
\label{eq:noise_wl}
\end{equation}
where $\sigma_\gamma$ is the intrinsic galaxy shear. In this paper, we assume
the intrinsic galaxy shear to be $\sigma_\gamma =0.3$ as the empirically
derived value \cite{Bernstein02} for all surveys.


\begin{widetext}

\begin{table}[th]
\begin{ruledtabular}
\begin{tabular}{cccrc}
survey & $z_{\rm m}$ & ${\bar n}_{\rm g}~[{\rm arcmin}^{-2}]$ & survey area \\
\hline
DES  & 0.8 & 10 &  5,000 ${\rm deg}^2$ & ($f_{\rm sky} \simeq 0.12$) \\
HSC  & 1.0 & 30 &  2,000 ${\rm deg}^2$ & ($f_{\rm sky} \simeq 0.05$) \\
LSST & 1.2 & 50 & 20,000 ${\rm deg}^2$ & ($f_{\rm sky} \simeq 0.50$) 
\end{tabular}
\end{ruledtabular}
\caption{Survey parameters adopted in this paper for DES, HSC and LSST. The
  parameter $z_{\rm m}$ denotes the mean redshift. The parameter ${\bar
    n}_{\rm g}$ denotes the mean number density of source galaxies at all
  redshifts, where the source galaxies are divided into five redshift bins in
  all surveys. The ranges of redshift are summarized in
  Table~\ref{tb:z-range}.}
\label{tb:survey_params}
\end{table}

\begin{table}[ht]
\begin{ruledtabular}
\begin{tabular}{cccccc}
$i$-th bin & 
1st & 2nd & 3rd & 4th & 5th \\
\hline
redshift range 
 & $0 < z \leq \frac{2}{5} z_{\rm m}$ 
 & $\frac{2}{5} z_{\rm m} < z \leq \frac{4}{5} z_{\rm m}$ 
 & $\frac{4}{5} z_{\rm m} < z \leq \frac{6}{5} z_{\rm m}$ 
 & $\frac{6}{5} z_{\rm m} < z \leq \frac{8}{5} z_{\rm m}$ 
 & $\frac{8}{5} z_{\rm m} < z \leq  z_{\rm max} $ 
\end{tabular}
\end{ruledtabular}
 \caption{The redshift ranges of the $i$-th bin in the case of dividing into
   five redshift bins. $z_{\rm m}$ is the mean redshift for each survey shown
   in Table~\ref{tb:survey_params}.  We assume five redshift bins and adopt
   these redshift ranges for all surveys.  $z_{\rm max}$ is the limitation of
   observations and we assume $z_{\rm max}=4.0$ for all surveys in this
   paper.}
 \label{tb:z-range}
\end{table}


\begin{figure}[t]
  \includegraphics[clip,keepaspectratio=true,width=0.95
    \textwidth]{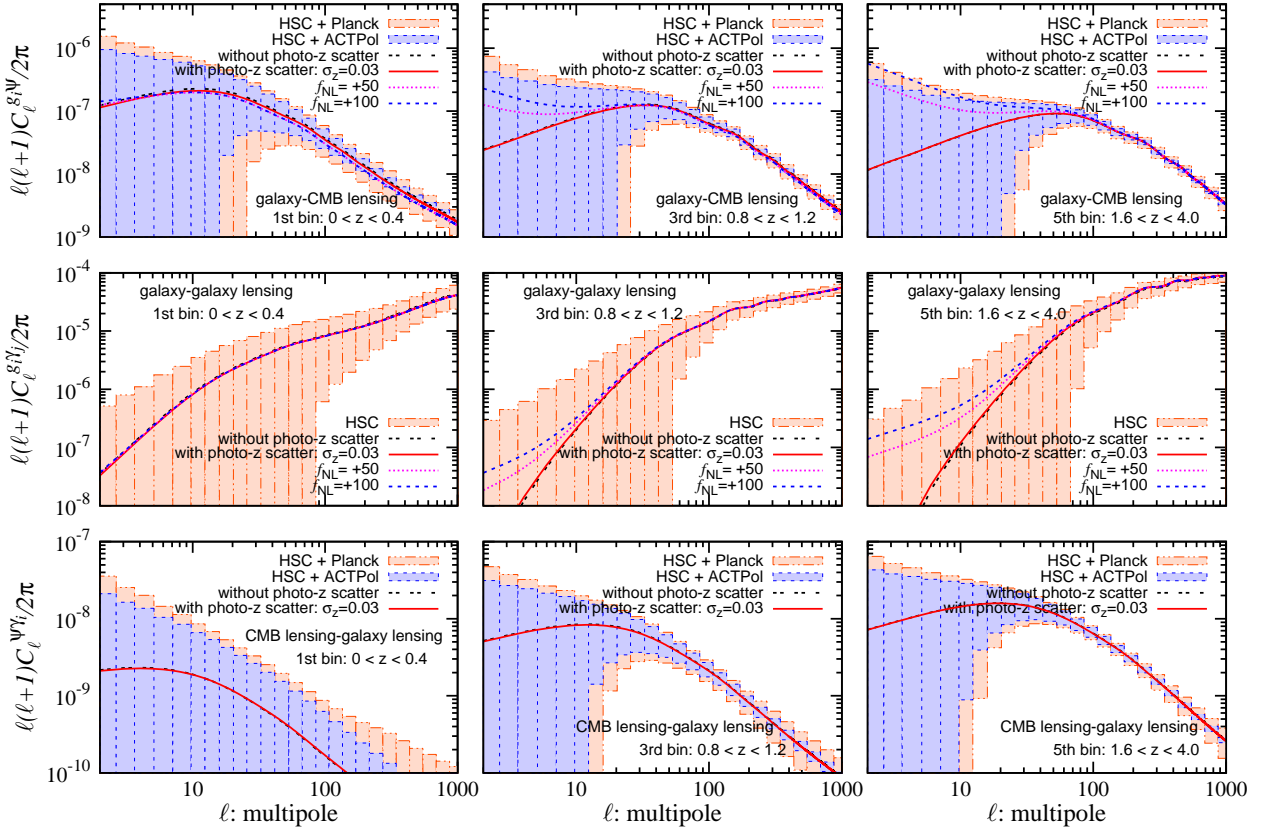}
  \caption{The angular power spectra for galaxy-CMB lensing (${\rm g}_i\psi$;
    Top), galaxy-galaxy lensing (${\rm g}_i\gamma_j$; Middle), and CMB
    lensing-galaxy lensing ($\psi \gamma_i$; Bottom). Here the subscripts
    $i,~j$ represent the redshift bin and we divided into 5 redshift bins for
    galaxy distribution~(g) and cosmic shear~($\gamma$). The angular power
    spectra shown here are $i=j=1,3,5$ from left to right, respectively. The
    boxed bars are the errors of the power spectra assuming HSC, Planck and/or
    ACTPol, and we adopt the logarithmic binning for illustrative purpose. We
    also plot the angular power spectra in the non-Gaussian case, $f_{\rm
      NL}=50,100$ and $M_{\rm obs} [h^{-1}M_\odot] =10^{11.2}, 10^{12.2}$. For
    comparison, we plot the contribution of the photo-$z$ error,
    $\sigma_z^{(i)}=0.03$ for each bin.}
\label{fig:CC_power}
\end{figure}

\end{widetext}


\subsection{Cross correlation angular power spectrum}

The effect of the primordial non-Gaussianity stands out the most in the
galaxy-galaxy auto correlation (gg) through the bias parameter. As many
previous works show, almost all the constraints on the primordial
non-Gaussianity come from the galaxy-galaxy power spectrum
\cite{Slosar08,Afshordi08,Xia10a,Xia10b,Xia11}. The cross correlation power
spectra, e.g., CMB Temperature-galaxy cross correlation ($T$g), however,
provide extra information and improve the errors of the parameters.

In our previous analysis \cite{Takeuchi10}, we paid a particular attention to
the CMB lensing-galaxy cross correlation ($\psi$g), and estimated the
contribution to the constraint on the primordial non-Gaussianity. Here, we
improve our analysis by adding the galaxy lensing shear, which may also
correlated with galaxy distribution and CMB lensing, and estimate the errors
of determining cosmological parameters including any possible auto and
cross correlations.

Fig.~\ref{fig:CC_power} shows an example of the cross correlation angular
power spectra, for galaxy-galaxy lensing (g$\gamma$), galaxy-CMB lensing
(g$\psi$) and CMB lensing-galaxy lensing ($\psi\gamma$). The boxes around the
curve show the expected measurement errors for HSC-$like$ survey (for an
illustrating purpose, we take the logarithmic binning, as $\Delta \ell \simeq
0.23 \ell$). For the cross correlations with CMB lensing, we show the errors
for two cases that one is for Planck and the other is Planck with ACTPol.


\section{Survey design} \label{sec:survey}

The errors in determining cosmological parameters from observations of LSS
depend a great deal on the survey design, such as the survey region, the mean
redshift and photometric redshift errors. Here we summarize the survey design
employed in our analysis.

\subsection{Redshift distribution of galaxies}

We assume the redshift distribution of galaxy samples with a function of the
form:
\begin{equation}
  n(z) = \bar{n}_{\rm g} 
  \frac{\beta}{\Gamma [(\alpha+1)/\beta]} 
  \frac{z^\alpha }{z_0^{\alpha+1}} 
  \exp \left[ -\left( \frac{z}{z_0} \right)^\beta \right] ,
\label{eq:dndz}
\end{equation}
where $\bar{n}_{\rm g}$ is the mean number density of source galaxies at all
redshifts, $\alpha$, $\beta$ and $z_0$ are free parameters.  We adopt
$\alpha=2.0$, $\beta=1.5$, and the parameter $z_0$ is related to the mean
redshift $z_{\rm m}$ as
\begin{equation}
  z_{\rm m} 
  = \int dz z \frac{n(z)}{\bar{n}_{\rm g}} 
  = z_0 \frac{\Gamma[(\alpha+2)/\beta]}{\Gamma[(\alpha+1)/\beta]},
\end{equation}
and $z_0$ is determined in such a way that the mean redshift $z_{\rm m}$ fits
the value in Table~\ref{tb:survey_params} for each survey. The normalization
of the redshift distribution function is fixed by the total number density of
galaxies:
\begin{equation}
  n^{\rm A} = \int_0^\infty dz n(z) .
\label{eq:total_ng}
\end{equation}


\subsection{Photometric redshift systematics}

In the galaxy imaging survey whose observables are galaxy distribution and
galaxy shear, we measure a large number of galaxies so that the systematic
uncertainties play a correspondingly important role with statistical
errors. Although systematics may include many effects and these effects depend
on the details of the observations \cite{Bernstein09,Das11}, we include only
the effect of photometric redshift errors.

Imperfect calibration of photometric redshifts induces a residual scatter and
bias with respect to the true redshifts, and uncertainties in the redshifts
distort the volume element.  Following the photometric redshift model as
described in \cite{Ma06}, the probability distribution of photometric redshift
$z_{\rm ph}$ given the true redshift $z$, $p(z_{\rm ph} | z)$ is modeled as a
Gaussian distribution
\begin{equation}
  p(z_{\rm ph} | z) = \frac{1}{\sqrt{2\pi}\sigma_z} 
  \exp \left[ - \frac{(z-z_{\rm ph} - z_{\rm bias})^2}
    {2\sigma_{z}^{2}} \right] ,
\end{equation}
where $\sigma_z(z)$ and $z_{\rm bias}(z)$ are the scatter and bias,
respectively, which are arbitrary function of redshift $z$.

The best-estimated distribution for objects in the $i$-th photometric redshift
bin with $z_{\rm ph}^{(i)} < z < z_{\rm ph}^{(i+1)}$, $n_i(z)$, can be written
as
\begin{eqnarray}
  n_i(z) 
  &=& \int_{z_{\rm ph}^{(i)}}^{z_{\rm ph}^{(i+1)}}
  dz_{\rm ph} n(z) p(z_{\rm ph}|z) , \nn \\
  &=& \frac{1}{2}n(z)\left[ {\rm erf}(x_{i+1}) 
    - {\rm erf}(x_{i}) \right] ,
\end{eqnarray}
where $x_i$ is defined as
\begin{equation}
  x_i \equiv 
  \left( z_{\rm ph}^{(i)} -z + z_{\rm bias}(z) 
  \right) / \sqrt{2}\sigma_z(z) .
\end{equation}
In the same way as Eq.~(\ref{eq:total_ng}), the total number density of
galaxies in the $i$-th bin becomes
\begin{equation}
  n_i^{\rm A} = \int_0^{\infty} dz n_i(z) .
\end{equation}

We assume a fiducial redshift scatter and bias following \cite{Das11} as
\begin{eqnarray}
  \label{eq:sys_scatter}
  \sigma_z(z) = \sigma_{z}^{(i)} (1+z) , \\
  \label{eq:sys_bias}
  z_{\rm bias}(z) = z_{{\rm bias}}^{(i)} (1+z) ,
\end{eqnarray}
where $\sigma_{z}^{(i)}$ and $z_{\rm bias}^{(i)}$ are defined for each
redshift bin.  We set the fiducial values as $\sigma_{z}^{(i)}=0.03$, $z_{\rm
  bias}^{(i)}=0$ for each redshift bin and all surveys.


\section{Forecasts} \label{sec:forecasts}

We estimate the parameter errors using the Fisher matrix formalism. We
summarize the Fisher matrix formalism taking into account the cross
correlation between CMB and LSS, and we describe the setup for the CMB and LSS
observables.


\subsection{Fisher matrix formalism} \label{sec:fisher}

Under the assumptions that the observables follow the Gaussian statistics, we
can obtain the information on a set of parameters ${\bm p}$ from the Fisher
matrix written as \cite{Tegmark97}
\begin{equation}
  F_{\ell, ij} = f_{\rm sky} \sum_{\l_{\rm min}}^{\l_{\rm max}} 
  \frac{(2\l +1)}{2}{\rm Tr} \left[ 
    {\bm C}_{\l,i} {\bm C}_\l^{-1} {\bm C}_{\l,j} {\bm C}_\l^{-1} 
    \right] ,
\end{equation}
where $f_{\rm sky}$ is the sky coverage of the experiment, ${\bm C}_\l$ is the
signal plus noise covariance matrix, and ${\bm C}_{\l ,i} $ is the derivative of
${\bm C}$ with respect to the parameter $p_i$.

For CMB experiments, we consider two cases: that one is the constraint from
Planck and the other is from Planck with ACTPol. It is well known that
combining the results of the satellite CMB experiment with the ground based one can
improve the constraints, e.g., combining WMAP with ACT\cite{Dunkley10},
ACBAR\cite{Reichardt09} or so on. For these two cases, we define the total
Fisher matrix as below, respectively, under the assumption for the size of
survey area as $f_{\rm sky}^{\rm ACTPol} < f_{\rm sky}^{\rm LSS} < f_{\rm
  sky}^{\rm Planck}$. Here $f_{\rm sky}^{\rm ACTPol}$, $f_{\rm sky}^{\rm LSS}$
and $f_{\rm sky}^{\rm Planck}$ are the sky coverage of the ACTPol, the LSS and
the Planck surveys, respectively, and we define the overlap region between CMB
experiments and LSS surveys as $f_{\rm sky}^{\rm Cross}$.


\begin{itemize}
\item{\bf LSS + Planck:}
\begin{multline}  
  F_{\ell, ij} = 
  f_{\rm sky}^{\rm Cross} 
    \sum_{\l_{\rm min}=2}^{\l_{\rm max}^{\rm LSS}} 
        {\cal F}_{\ell, ij}^{\rm Cross} 
  + f_{\rm sky}^{\rm Planck} \sum_{\l_{\rm max}^{\rm LSS}+1}^{\l_{\rm max}^{\rm
      CMB}}    {\cal F}_{\ell, ij}^{\rm CMB} \\
  + (f_{\rm sky}^{\rm Planck} - f_{\rm sky}^{\rm Cross}) \sum_{\l_{\rm
      min}=2}^{\l_{\rm max}^{\rm LSS}} {\cal F}_{\ell, ij}^{\rm CMB} ,
\label{eq:Fisher_I}
\end{multline}
\end{itemize}

\begin{itemize}
\item{\bf LSS + Planck + ACTPol:} 
\begin{multline}  
  F_{\ell, ij} = 
  f_{\rm sky}^{\rm Cross} 
  \sum_{\l_{\rm min}=2}^{\l_{\rm max}^{\rm LSS}} 
 \left. {\cal F}_{\ell, ij}^{\rm Cross} \right|_{\rm ACTPol} \\
  + f_{\rm sky}^{\rm ACTPol}
  \sum_{\l_{\rm max}^{\rm LSS}+1}^{\l_{\rm max}^{\rm CMB}} 
  \left. {\cal F}_{\ell, ij}^{\rm CMB} \right|_{\rm ACTPol} \\
  + (f_{\rm sky}^{\rm LSS} - f_{\rm sky}^{\rm Cross}) 
  \sum_{\l_{\rm
      min}=2}^{\l_{\rm max}^{\rm LSS}} 
  \left. {\cal F}_{\ell, ij}^{\rm Cross} \right|_{\rm Planck} \\
  + (f_{\rm sky}^{\rm Planck} - f_{\rm sky}^{\rm LSS}) 
  \sum_{\l_{\rm min}=2}^{\l_{\rm max}^{\rm
      LSS}} 
  \left. {\cal F}_{\ell, ij}^{\rm CMB} \right|_{\rm Planck}\\
  + (f_{\rm sky}^{\rm Planck} - f_{\rm sky}^{\rm ACTPol}) 
  \sum_{\l_{\rm max}^{\rm LSS}+1}^{\l_{\rm max}^{\rm CMB}} 
  \left. {\cal F}_{\ell, ij}^{\rm CMB} \right|_{\rm Planck} , \\ 
\label{eq:Fisher_II}
\end{multline}
\end{itemize}
where
\begin{eqnarray}
  {\cal F}_{\ell, ij}^{\rm Cross} &=& \frac{(2\l +1)}{2}{\rm Tr} 
  \left[ 
    {\bm C}_{\l,i}^{\rm Cross} ({\bm C}_\l^{\rm Cross})^{-1} 
    {\bm C}_{\l,j}^{\rm Cross} ({\bm C}_\l^{\rm Cross})^{-1} 
    \right] , \nn \\\\
  {\cal F}_{\ell, ij}^{\rm CMB} &=& \frac{(2\l +1)}{2}{\rm Tr} \left[ 
    {\bm C}_{\l,i}^{\rm CMB} ({\bm C}_\l^{\rm CMB})^{-1} 
    {\bm C}_{\l,j}^{\rm CMB} ({\bm C}_\l^{\rm CMB})^{-1} \right] , \nn \\
\end{eqnarray}
and different superscripts for $\l_{\rm max}$ and $f_{\rm sky}$ represent
different observations or CMB experiments, e.g., $\ell_{\rm max}^{\rm CMB}$
and $\ell_{\rm max}^{\rm CMB}$ are the maximum multipole we can take for the
CMB experiment and the LSS survey, respectively. Here $\ell_{\rm min}$ is the
minimum multipole we can take and we use $\ell_{\rm min}=2$ in all cases. We
summarize these values in Table \ref{tb:survey_HSC}.


\begin{table}[t]
\begin{ruledtabular}
  \begin{tabular}{c||c|c|c}
\multicolumn{1}{c}{}    & 
\multicolumn{1}{c}{CMB} & 
\multicolumn{1}{c}{LSS} & 
\multicolumn{1}{c}{Cross: [CMB$\times$LSS]} \\
\hline \hline 
$C_\l^{\rm XY}$ 
  & $TT$,~$EE$,~$\psi \psi$,~$TE$,~$T\psi$ 
  & gg,~$\gamma \gamma$,~g$\gamma$ 
  & $T$g,~$T\gamma$,~$\psi$g,~$\psi \gamma$ \\
$\l_{\rm max}$ & 3000 (8000)       & 800   & 800            \\
$f_{\rm sky}$ & 0.65 (0.05) & 0.05  & 0.05            \\
  \end{tabular}
\end{ruledtabular}
\caption{Survey parameters for our analysis. We assume Planck (ACTPol) for CMB
  experiments or their combination. For the LSS survey, we assume the HSC
  survey and the survey area of HSC fully overlaps with Planck or ACTPol
  experiments.}
\label{tb:survey_HSC}
\end{table}


The covariance matrix ${\bm C}_\l$ and each of its matrix elements is 
defined as
\begin{eqnarray}
  [{\bm C}_\l^{\rm Cross, CMB}]_{\rm XY} 
  &=& C_\l^{\rm XY} + N_\l^{\rm XY}\delta_{\rm XY} , \\
  {\rm where}~X,Y &=& 
\begin{cases}
  T,E,\psi,g_{i},\gamma_{i} & ({\rm for~Cross})\\
  T,E,\psi & ({\rm for~CMB})
\end{cases} , \nn
\end{eqnarray}
$C_\ell^{\rm XY}$ is the angular power spectrum and $N_\ell^{\rm XY}$ is the noise
spectrum for each auto or cross correlation.  We defined $N_\ell^{TT}$ and
$N_\ell^{EE}$ as $N_\ell^{T}$ and $N_\ell^{P}$ in Eq.~(\ref{eq:noise_cmb}),
$N_{\ell}^{{\rm g}_i{\rm g}_j}$ and $N_\ell^{\gamma_i\gamma_j}$ in
Eqs.~(\ref{eq:noise_gal}) and (\ref{eq:noise_wl}), respectively, and
$N_\ell^{\psi\psi}$ is the noise spectrum of CMB lensing potential we estimate
following the technique developed in \cite{Okamoto03}. Then we assume that
there is no correlation of noises between the different observables and
different redshift bins.  Moreover, we assume that $E$-mode polarization ($E$)
has correlation only with temperature anisotropies ($T$) because $E$-mode
polarization can be generated through the Thomson scattering dominantly on the
last scattering surface, $C_\ell^{E\psi} = C_\ell^{E{\rm g}_i} =
C_\ell^{E{\gamma}_i} =0$.

Incidentally the assumption that $C_\ell^{E\psi} = 0$ is not strictly correct
\cite{Lewis11} because the large-scale polarization $E$-mode generated by
scattering at reionization correlates with sources of CMB lensing potential
over long distance. Moreover, if the galaxies are at high-redshift, the
assumption that $C_\ell^{E{\rm g}_i} = 0$ is also not correct
\cite{Challinor11}.  However the influence of these assumptions for our
results, especially constraints on the primordial non-Gaussianity, can be
small because sample galaxies in our analysis are at low redshift.

Here we explain the detail of our Fisher matrices given in
Eqs.~(\ref{eq:Fisher_I}) and (\ref{eq:Fisher_II}). In Eq.~(\ref{eq:Fisher_I})
the first term is responsible for the cross correlation between LSS and CMB,
the second one is for CMB observables at the higher multipoles than the LSS
survey can probe, and the last one is for CMB observables in the remaining
survey area of CMB which does not overlap with the LSS survey.  In
Eq.~(\ref{eq:Fisher_II}) the first term is responsible for the cross
correlation between LSS and CMB with ACTPol, the second one is for CMB
observables with ACTPol at the higher multipoles than the LSS survey can
probe, the third one is for the cross correlation between LSS and CMB with
Planck in the overlapping region which does not overlap with ACTPol, the forth
one is for CMB observables with Planck in the region which does not overlap
with the LSS survey, and the last one is for CMB observables with Planck at
the higher multipoles than the LSS survey can probe in the region which does
not overlap with ACTPol.

In our analysis, the total number of non-zero angular power spectra is 81 for
the full cross correlation case ($T,E,\psi,{\rm g}_i,\gamma_i$), while it is 5
for the only CMB case ($T,E,\psi$).


\subsection{Setup}\label{sec:setup}

Given the measurement design of the surveys, we can estimate the errors in
determining the cosmological parameters using the Fisher matrix formalism. The
formalism tells us how well the given surveys can measure the cosmological
parameters around fiducial cosmological model. Therefore, the parameter
forecasts by Fisher matrix formalism depend on the choice of the fiducial
model and the number of free parameters.

As for our fiducial cosmological model, we assume a flat $\Lambda${\rm CDM}
model and we include the following 9 cosmological parameters, which is based
on the WMAP seven-year results \cite{WMAP7}. The density parameters of cold
dark matter, baryon and dark energy are $\Omega_{\rm c}h^2$, $\Omega_{\rm
  b}h^2$ and $\Omega_{\rm \Lambda}$, respectively; dark energy equation of
state parameter is $w$; the optical depth at reionization is $\tau$; and the
primordial power spectrum parameters are the spectrum tilt $n_{\rm S}$ and the
amplitude of the primordial power spectrum $\Delta_{\cal R}^2$ which is
normalized at $k_0 =0.002~\rm{Mpc}^{-1}$. We also include the primordial
non-Gaussianity of local type modeled by the non-linear parameter $f_{\rm
  NL}$. For specifying the galaxy bias, we include the minimum mass of halo
hosting the galaxies, for which we can observe $M_{\rm obs}[h^{-1}M_\odot]$ as a
free parameter.  The fiducial values are
\begin{multline}
  \{ 100\Omega_{\rm b}h^2,~ \Omega_{\rm c}h^2,~ \Omega_{\Lambda},~ \tau,~
  n^{\rm S},~ \Delta_{\cal R}^{2},~ w,~ f_{\rm NL},~ \log M_{\rm obs} \} \\ 
  = \{ 0.20 ,~ 0.1109 ,~ 0.72 ,~ 0.086 ,~ 0.96 ,~ 2.4 \times 10^{-9},~ -1 ,~ 0
  ,~ 11.7 \}. \nn
\end{multline}
Throughout this paper we assume a spatially flat Universe, and the Hubble
parameter is adjusted to keep our Universe flat when we vary $\Omega_\Lambda$.

To calculate the angular power spectra $C_l$ including non-linear effects on
the angular power spectrum of the galaxy distribution, cosmic shear and
lensing potential, we use the CAMB code \cite{Lewis00} and the HALOFIT code
\cite{Smith03}. We, however, want to remove the burden of the uncertainty of
the non-linear calculation as much as possible, so our estimation preferably
is based on the information from large-scale for LSS where the linear
prediction is reliable.

In our estimation, we include the information from temperature anisotropies
($T$), $E$-mode polarization ($E$) and reconstructed lensing potential ($\psi$)
for CMB. The range of multipoles are $2 \le \l \le 3000$ and the survey area
is taken as $f_{\rm sky}=0.65$ for Planck and $f_{\rm sky}=0.05$ for ACTPol,
respectively. On the other hand, for the galaxy survey (${\rm g}$) and galaxy
weak lensing survey ($\gamma$), the ranges of multipole are $2 \le \l \le 800$
and the survey areas are $f_{\rm sky}=0.05$ for both of them with
HSC. Furthermore we take into account all the cross correlation we can
consider and we assume the optimal condition that there is no correlation between
different patches, so that the area having a correlation between CMB and LSS
corresponds to the overlapped survey area. We summarize the values we use in
the following calculation in TABLE~\ref{tb:survey_HSC}.


\section{Result} \label{sec:result} 

\subsection{Signal-to-Noise ratio}
 
To see the errors in the cross correlation power spectra more
quantitatively, we estimate the signal-to-noise ratio ($S/N$) for each
$\l$-bin and the cumulative ($S/N$).  We define the $(S/N)^2$ for a
cross correlation between X and Y, following \cite{Peiris+00}, as \\

{\bf $\bullet$ $S/N$ for each $\l$-bin:}
\begin{multline}
  \left( \frac{S}{N} \right)^2 (\l) 
  \equiv f_{\rm sky} 
  \sum_{\l_{\rm min}^{(i)}}^{\l_{\rm max}^{(i)}}(2\l+1)  \\
  \times \frac{(C^{\rm XY}_\l )^2}
  {(C^{\rm XX}_\l+N^{\rm XX}_\l)(C^{\rm YY}_\l+N^{\rm YY}_\l) 
    + (C^{\rm XY}_\l)^2} , 
\label{eq:sn_l}
\end{multline}

{\bf $\bullet$ stacked $S/N$:}
\begin{multline}
  \left( \frac{S}{N} \right)^2 (\leq \l_{\rm max}) \equiv 
  f_{\rm sky} \sum_{\l=2}^{\l_{\rm max}} (2\l +1) \\
  \times \frac{(C^{\rm XY}_\l )^2}
  {(C^{\rm XX}_\l+N^{\rm XX}_\l)(C^{\rm YY}_\l+N^{\rm YY}_\l) 
    + (C^{\rm XY}_\l)^2} , 
\label{eq:cum_sn}
\end{multline}
where $\l_{\rm max}^{(i)}$ and $\l_{\rm min}^{(i)}$ are the maximum and
minimum multipoles of the $i$-th multipole bin, respectively, $C_\l^{\rm XY}$
and $N_\l^{\rm XX}$, $N_\l^{\rm YY}$ are the fiducial and noise spectra,
respectively. The index ${\rm X, Y}$ represents $\{{\rm X, Y}\} \in \{ T, E,
\psi, {\rm g}, \gamma\}$.

Figure ~\ref{fig:StoN} shows an example of $(S/N)^2$ for each $\l$ given by
Eq.~(\ref{eq:sn_l}) and cumulative $(S/N)^2$ by Eq.~(\ref{eq:cum_sn}), for the
galaxy-galaxy lensing (g$\gamma$), galaxy-CMB lensing (g$\psi$) and CMB
lensing-galaxy lensing ($\psi\gamma$) cross correlations.  For the galaxy-CMB
lensing and CMB lensing-galaxy lensing with HSC-Planck experiments, the
amplitude of $(S/N)^2$ decreases at high-$\l$ bins and the stacked $(S/N)^2$
saturates due to the noise of CMB lensing at small scales. On the other hand,
we can expect the lager $(S/N)^2$ at high-$\l$ bins for the case with
HSC-ACTPol.  The amplitude of $(S/N)^2$ at low-$\l$ bins where the signature
of the primordial non-Gaussianity shines out is very low in all the
cases. However, comparing galaxy-CMB lensing with galaxy-galaxy lensing
signals, the amplitude of the former is larger than the latter in
high-redshift bins and the difference becomes even greater for the with
HSC-ACTPol.


\begin{widetext}

\begin{figure}[t]
\includegraphics[clip,keepaspectratio=true,width=0.95
  \textwidth]{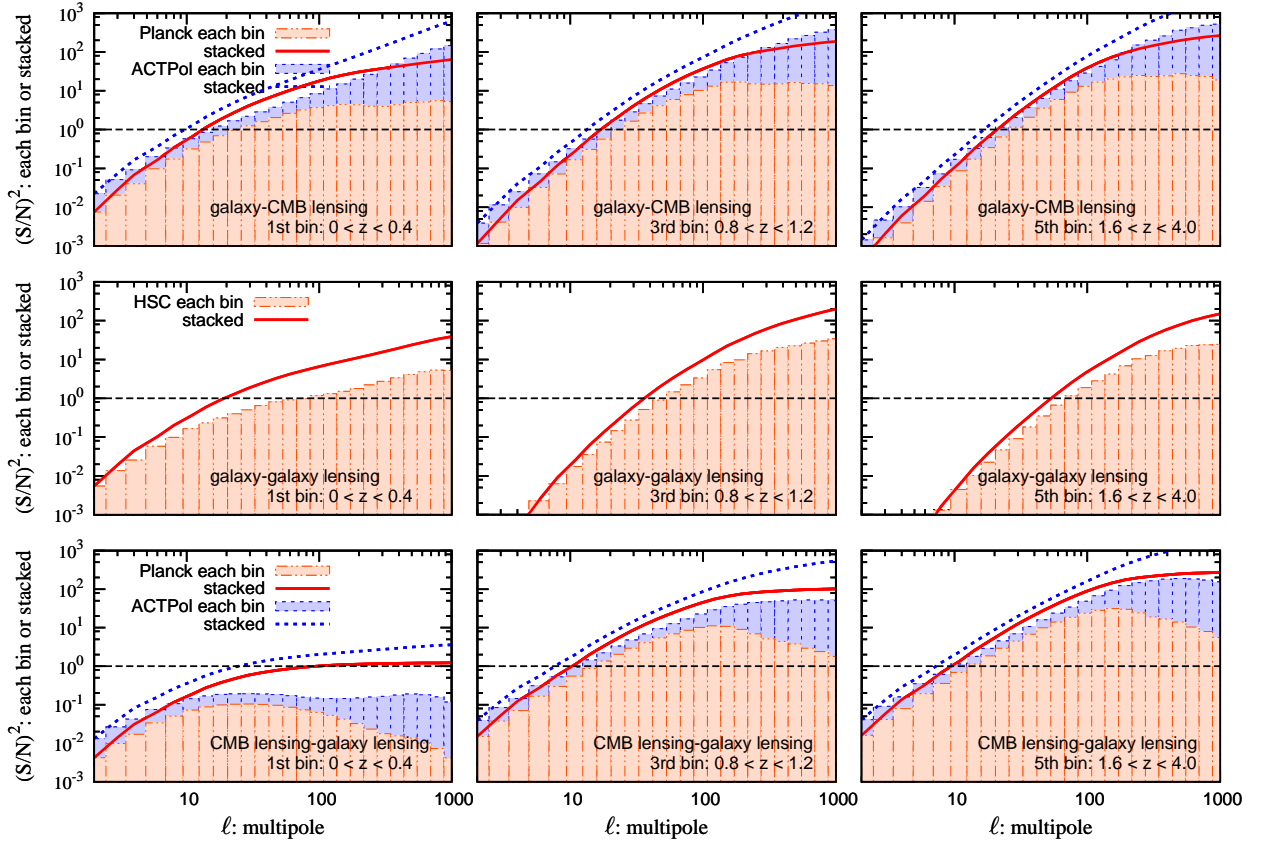}
\caption{Signal-to-Noise ratio for some angular power spectra. We show the $S/N$
  for each bin (box) and stacked $S/N$ (line), respectively.  Top panels are
  for galaxy-CMB lensing (${\rm g}_i\psi$) angular power spectra, Center
  ones are for galaxy-galaxy lensing (${\rm g}_i\gamma_j$), and Bottom
  ones are for CMB lensing-galaxy lensing ($\psi \gamma_j$) based on 
  the fiducial model of Fig.~\ref{fig:CC_power}.}
\label{fig:StoN}
\end{figure}


\begin{figure}[t]
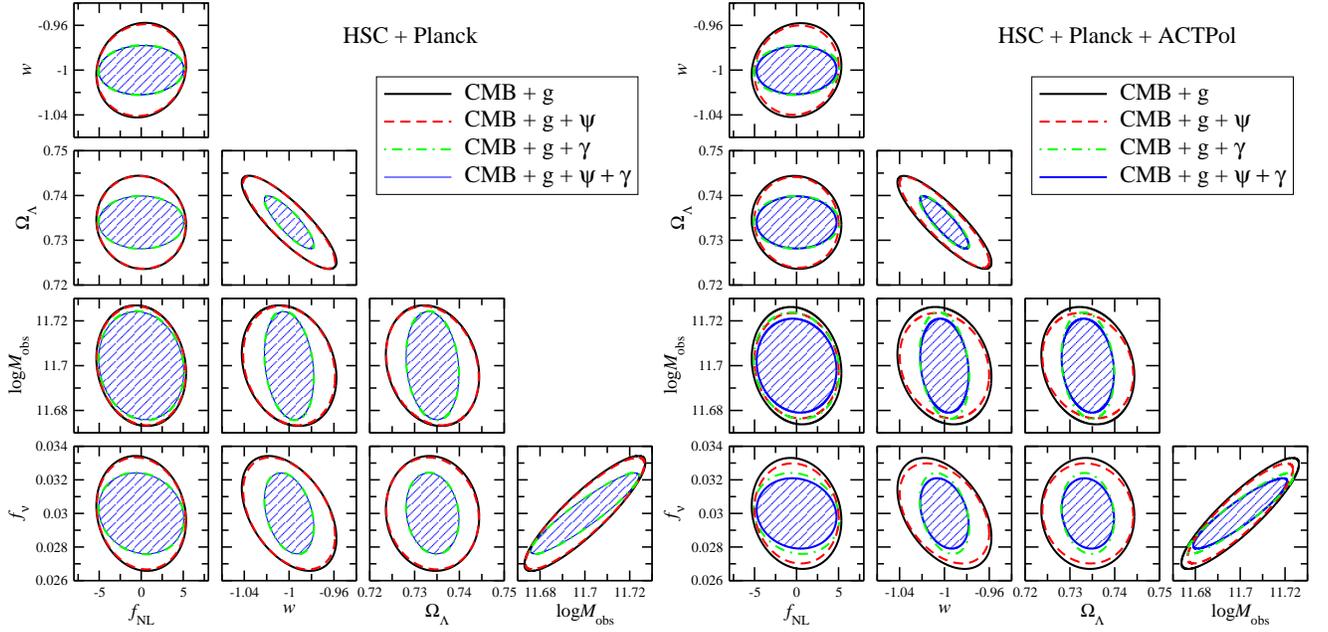

    \includegraphics[clip,keepaspectratio=true,width=0.48
      \textwidth]{fig4_lens-PLA.eps}
    \includegraphics[clip,keepaspectratio=true,width=0.48
      \textwidth]{fig4_lens-ACT.eps}
\caption{Projected 1$\sigma$(68\%) confidence constraints in some parameter
  spaces for the $f_\nu \neq 0$ model. We show contours from CMB and galaxy
  distribution (black thick); from CMB, galaxy distribution and CMB lensing
  (red dashed); from CMB, galaxy distribution and galaxy lensing (green
  dot-dashed); and from all the observables (blue thin). All auto and cross
  correlations between these observables are taken into account for the
  constraints. Here we assume the HSC survey and Planck experiment (Left) or HSC
  and combination of Planck and ACTPol (Right).}
\label{fig:cont_lens}
\end{figure}
\begin{table}[th]
\begin{ruledtabular}
\begin{tabular}{lcccccccccc}
& \makebox[3.6em][c]{$100\Omega_{\rm b}h^2$}
& \makebox[3.6em][c]{$\Omega_{\rm c}h^2$}
& \makebox[3.6em][c]{$\Omega_{\Lambda}$}
& \makebox[3.6em][c]{$\tau$}
& \makebox[3.6em][c]{$f_\nu$}
& \makebox[3.6em][c]{$n^{\rm S}$}
& \makebox[4.2em][c]{$\Delta_{\cal R}^{2} \times 10^{-9}$}
& \makebox[3.2em][c]{$w$} 
& \makebox[2.8em][c]{$f_{\rm NL}$} 
& \makebox[3.6em][c]{$\log M_{\rm obs}$} \\
\hline
galaxy clustering only ({\bf HSC}): $C_\ell^{\rm gg}$ & 
  0.94 & 0.026 & 0.011 & 0.11 & 0.029 & 0.13 & 0.49 & 0.053 & 5.9 & 0.60 \\ 
\hline \hline
{\bf HSC + Planck} \\ 
CMB only &
  0.015 & 0.0013 & 0.19 & 0.0050 & 0.020 & 0.0047 & 0.042 & 0.69 & --- & --- \\
CMB + $C_\ell^{\rm gg}$ & 
  0.011 & 0.00055 & 0.010 & 0.0044 & 0.0034 & 0.0027 & 0.029 & 0.042 & 5.4 &
  0.027 \\ 
\multicolumn{11}{c}{\hdashrule[0.5ex]{\textwidth}{0.1pt}{1mm}} \\
CMB + $C_\l^{\rm gg}$ + $C_\l^{Tg}$
  &
    0.011   &  
    0.00055 &  
    0.010   &  
    0.0044  &  
    0.0034  &  
    0.0027  &  
    0.029   &  
    0.042   &  
    5.3     &  
    0.027      \\
\hspace{5pt} + $C_\l^{\psi \psi}$ + $C_\l^{\psi g}$ + $C_\l^{T\psi}$
  &
    0.011   &  
    0.00053 &  
    0.010   &  
    0.0043  &  
    0.0033  &  
    0.0027  &  
    0.028   &  
    0.041   &  
    5.3     &  
    0.027      \\
\hspace{5pt} + $C_\l^{\gamma \gamma}$ + $C_\l^{\gamma g}$ + $C_\l^{T \gamma}$
  &
    0.011   &  
    0.00031 &  
    0.0059  &  
    0.0044  &  
    0.0024  &  
    0.0024  &  
    0.027   &  
    0.022   &  
    5.1     &  
    0.024      \\
\hspace{5pt} + $C_\l^{\psi \psi}$ + $C_\l^{\gamma \gamma}$ 
 + $C_\l^{\psi g}$ + $C_\l^{\gamma g}$ \\
\hspace{15pt} + $C_\l^{T\psi}$ + $C_\l^{T\gamma}$ 
 + $C_\l^{\psi \gamma}$
  &
    0.011   &  
    0.00031 &  
    0.0059  &  
    0.0043  &  
    0.0024  &  
    0.0023  &  
    0.026   &  
    0.022   &  
    5.1     &  
    0.024      \\
\hline \hline
{\bf HSC + Planck + ACTPol} \\
CMB only & 
  0.0076 & 0.0010 & 0.18 & 0.0046 & 0.018 & 0.0038 & 0.036 & 0.66 & --- & --- \\
CMB + $C_\ell^{\rm gg}$ & 
  0.0076 & 0.00052 & 0.010 & 0.0041 & 0.0033 & 0.0025 & 0.028 & 0.042 & 5.4 &
  0.026 \\ 
\multicolumn{11}{c}{\hdashrule[0.5ex]{\textwidth}{0.1pt}{1mm}} \\
CMB + $C_\l^{\rm gg}$ + $C_\l^{Tg}$ 
  &
    0.0076  &  
    0.00052 &  
    0.010   &  
    0.0041  &  
    0.0033  &  
    0.0025  &  
    0.028   &  
    0.042   &  
    5.3     &  
    0.026      \\
\hspace{5pt} + $C_\l^{\psi \psi}$ + $C_\l^{\psi g}$ + $C_\l^{T\psi}$
  &
    0.0076  &  
    0.00050 &  
    0.010   &  
    0.0039  &  
    0.0030  &  
    0.0024  &  
    0.026   &  
    0.040   &  
    5.0     &  
    0.024      \\
\hspace{5pt} + $C_\l^{\gamma \gamma}$ + $C_\l^{\gamma g}$ + $C_\l^{T \gamma}$
  &
    0.0074  &  
    0.00030 &  
    0.0059  &  
    0.0041  &  
    0.0024  &  
    0.0022  &  
    0.025   &  
    0.022   &  
    5.1     &  
    0.024      \\
\hspace{5pt} + $C_\l^{\psi \psi}$ + $C_\l^{\gamma \gamma}$ 
 + $C_\l^{\psi g}$ + $C_\l^{\gamma g}$ \\
\hspace{15pt} + $C_\l^{T\psi}$ + $C_\l^{T\gamma}$ 
 + $C_\l^{\psi \gamma}$
  &
    0.0073  &  
    0.00029 &  
    0.0058  &  
    0.0039  &  
    0.0021  &  
    0.0022  &  
    0.024   &  
    0.022   &  
    4.8     &  
    0.021      
\end{tabular}
\end{ruledtabular} 
\caption{Expected marginalized error (1$\sigma$) and these
  forecasts correspond to the results of Fig.~\ref{fig:cont_lens}.}
\label{tb:lens}
\end{table}

\end{widetext}


\begin{figure}[t]
  \includegraphics[clip,keepaspectratio=true,width=0.4
      \textwidth]{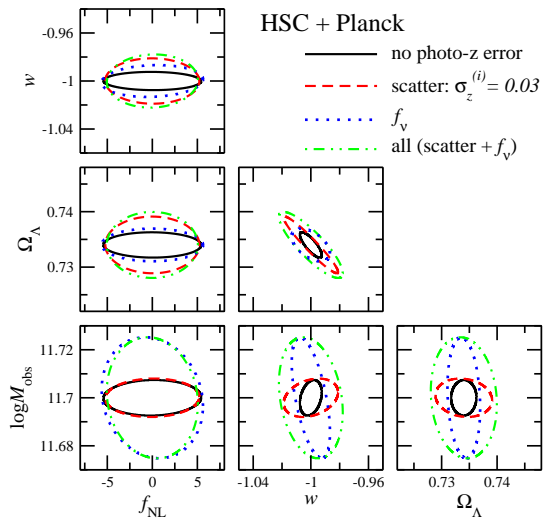}
  \caption{Projected 1$\sigma$(68\%) confidence constraints in some parameter
    spaces from Planck and HSC experiments including auto and cross
    correlations between CMB, CMB lensing, galaxy distribution and galaxy
    lensing. We show the contours with or without the photometric redshift
    scatter for the $\Lambda$CDM model, the contours without the photometric
    redshift scatter for the $f_\nu \neq 0$ model 
    and the contour with the photometric redshift scatter for the $f_\nu = 0$
    model.}
\label{fig:cont_fid}
\end{figure}


\subsection{Constraints on the primordial non-Gaussianity}

We estimate the errors in the determination of cosmological parameters for
the HSC-$like$ survey using the Fisher matrix formalism, and show the main
results, namely, the 1$\sigma$ confidence limit constraints in
Fig.~\ref{fig:cont_lens} and Table~\ref{tb:lens}, in which we consider 10
cosmological parameters: the fraction of the massive neutrino density
parameter to the matter density parameter $f_\nu(=0.03)=\Omega_\nu/\Omega_{\rm
  m}$ in addition to the nine cosmological parameters defined in
Sec.~\ref{sec:setup}.

The simplest case, which is from only galaxy clustering, puts the constraint on
$\fnl$ as $\Delta \fnl \sim 5.9$, and the constraint can be improved to
$\Delta \fnl \sim 5.4$ by combining the CMB-only result though CMB has no
information about primordial non-Gaussianity on the level of the 2-point function
without cross correlations with galaxy clustering.  This is because the
information from CMB helps to determine the galaxy bias by breaking
degeneracies with other cosmological parameters, and then the degeneracy
between $\fnl$ and $M_{\rm obs}$ can be broken effectively. Hence the
constraint on $M_{\rm obs}$ is also improved dramatically after combining CMB
information, which can be seen in Table~\ref{tb:lens}, and these tendencies
have been seen in our previous analysis \cite{Takeuchi10}.

We show four cases to see the improvement by adding the different observables
in Fig.~\ref{fig:cont_lens}.  In the first case we use CMB and galaxy
distribution (CMB + g); in the second case we add CMB lensing to the first
case (CMB + g + $\psi$); in the third case we add shear to the first case (CMB
+ g + $\gamma$); and in the last case we use all the information available.
We take into account all the available power spectra.  The selection of auto
and cross correlations in Fig.~\ref{fig:cont_lens} and Table~\ref{tb:lens}
corresponds to each other in each case.  We obtain the constraints on $f_{\rm
  NL}$ as $\Delta f_{\rm NL} \sim 5.1$ with Planck and HSC. Moreover, we can
improve the constraint by including ACTPol as $\Delta f_{\rm NL} \sim 4.8$,
thanks to the galaxy-CMB lensing cross correlations.


\section{Discussion} \label{sec:discussion}

\subsection{Selection of the fiducial model}

We assume imaging surveys for the LSS galaxy surveys, and we have to take into
account the influence of the various systematics, e.g.,, magnification
effects due to gravitational lensing \cite{Namikawa11}, photometric
redshift(photo-$z$) errors \cite{Ma06} and so on (see, for example,
\cite{Bernstein09,Das11}).  Moreover, as mentioned in the previous section,
the parameter forecasts by Fisher matrix formalism depend on the choice of the
fiducial model and free parameters. In Fig.~\ref{fig:cont_fid}, we summarize
the effects of selecting fiducial models.


\subsubsection{Photometric redshift error}

For the constraint on $f_{\rm NL}$, any significant difference is not found
between with and without the photometric redshift scatter systematics. This is
because the redshift scatter changes the overall amplitude of the galaxy
distribution (g) and cosmic shear ($\gamma$) as shown in
Fig.~\ref{fig:CC_power}. On the other hand, the effect of non-Gaussianity,
$f_{\rm NL}$, emerges only at large angular scales, and it can be 
distinguishable.  However, the effects of the photometric redshift error are
significant for the parameters which are determined through the information
about the redshift evolution such as $\Omega_\Lambda$ and $w$. Although the
effect of the parameter $M_{\rm obs}$ is to change the bias parameter and is
similar to the effect of the redshift scatter, a significant difference is not
found in the figure.


\subsubsection{Massive neutrino}

Various experiments, such as ground based experiments of neutrino oscillation,
predict the non-zero neutrino mass \cite{Schwetz08,Maltoni04}. Therefore, we
need to include the mass of neutrinos for a more accurate estimation of the
cosmological parameters. There are, however, some debatable subtle arguments
about taking the mass of neutrinos into account in models for the mass
function and the halo bias.  These models have been tested by N-body
simulations without massive neutrinos and are not guaranteed enough for
cosmological models with massive neutrinos. The analytic formula for the
scale-dependent bias presented in Eq.~(\ref{eq:deltabk}) is also for models
without massive neutrinos.

For the mass function, it is found that the influence of massive
neutrinos is very small \cite{Ichiki11}.  Thus, we just assume that the
halo mass function and halo bias models can be applied even if the
massive neutrinos are included.

Comparing the results of $f_\nu = 0$ and $f_\nu \neq 0$ models in
Fig.~\ref{fig:cont_fid}, a significant difference can be seen in the
constraint on $M_{\rm obs}$.  The existence of massive neutrinos affects the
large-scale structure formation through its large velocity dispersion and
alter the amplitude of the matter power spectrum. This effect is similar to
the effect from $M_{\rm obs}$ and these two parameters are strongly
degenerated with each other as shown in Fig.~\ref{fig:cont_lens}.  Note that
the degrade of the constraint on $M_{\rm obs}$ in the case of the $f_\nu \neq
0$ model comes from the strong degeneracy between them and we have confirmed
that there is little change in the constraint if we fix the value of $f_\nu$
(and do not take it as a free parameter).  For the constraints on $f_{\rm
  NL}$ itself, however, the difference of the fiducial models whether $f_\nu
=0$ or not is not so significant.

In summary, focusing on the constraint of $f_{\rm NL}$, the selection of the
fiducial model makes little impact on the result.  On the other hand, the
constraints on some other parameters depend on the fiducial model, especially
on the existence of neutrino mass and photometric redshift error.


\begin{widetext}
 
\begin{figure}[t]
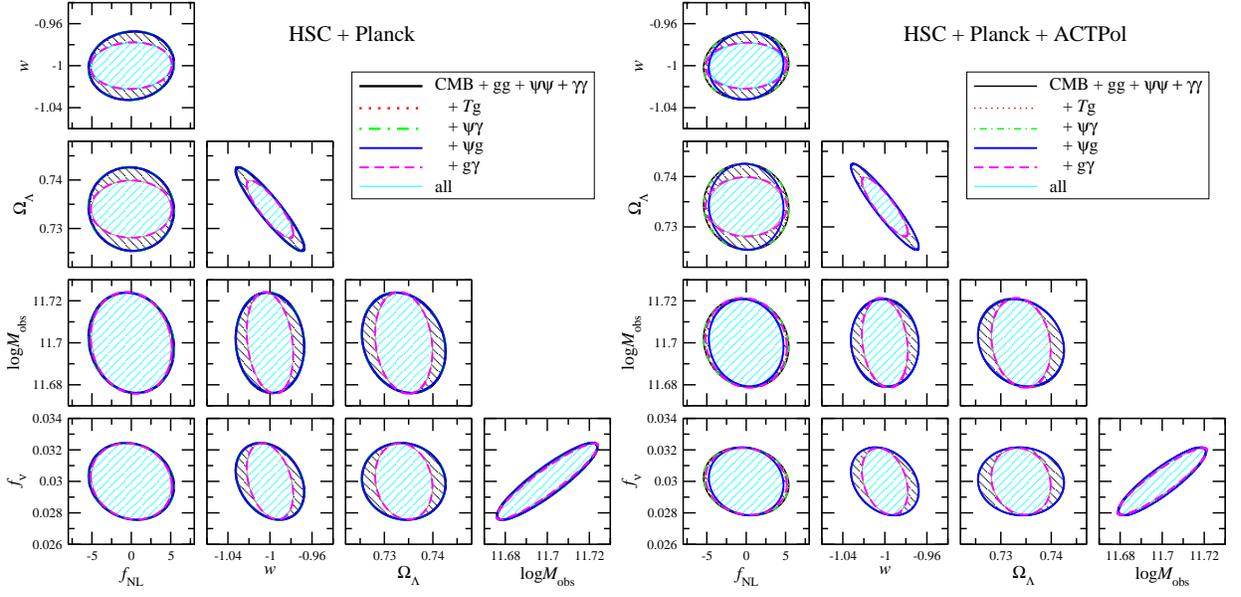

    \includegraphics[clip,keepaspectratio=true,width=0.45
      \textwidth]{fig6_cross-PLA.eps}
    \includegraphics[clip,keepaspectratio=true,width=0.45
      \textwidth]{fig6_cross-ACT.eps}
\caption{ Same as Fig.~\ref{fig:cont_lens} but we highlight the contribution
  from the cross correlations to the constraints.  The base constraints are
  from CMB($TT$,$EE$,$TE$), galaxy distribution (gg), CMB lensing ($\psi
  \psi$) and galaxy lensing ($\gamma \gamma$) auto correlations (black thin
  line).  The other contours show improvements by including cross correlation
  as indicated in the figure.  Here we consider the HSC survey and Planck with
  and without ACTPol experiment (left and right panels, respectively).  }
\label{fig:cont_cross}
\end{figure}
\begin{table}[t]
\begin{ruledtabular}
\begin{tabular}{lcccccccccc}
 & \makebox[2.5em][c]{$100\Omega_{\rm b}h^2$}
& \makebox[2.5em][c]{$\Omega_{\rm c}h^2$}
& \makebox[2.5em][c]{$\Omega_{\Lambda}$}
& \makebox[2.5em][c]{$\tau$}
& \makebox[2.5em][c]{$f_\nu$}
& \makebox[2.5em][c]{$n^{\rm S}$}
& \makebox[2.5em][c]{$\Delta_{\cal R}^{2} \times 10^{-9}$}
& \makebox[2.5em][c]{$w$} 
& \makebox[2.5em][c]{$f_{\rm NL}$} 
& \makebox[2.5em][c]{$\log M_{\rm obs}$}\\
\hline
{\bf HSC + Planck} \\
CMB + $C_\l^{\rm gg}$ + $C_\l^{\psi \psi}$ + $C_\l^{\gamma \gamma}$
 &
        0.011     &  
        0.00034   &  
        0.0086    &  
        0.0043    &  
        0.0024    &  
        0.0024    &  
        0.026     &  
        0.033     &  
        5.4       &  
        0.024        \\
\hspace{5pt} + $C_\l^{Tg}$  
 &
        0.011     &  
        0.00034   &  
        0.0086    &  
        0.0043    &  
        0.0024    &  
        0.0024    &  
        0.026     &  
        0.033     &  
        5.3       &  
        0.024        \\
\hspace{5pt} + $C_\l^{\psi \gamma}$ 
 &
        0.011     &  
        0.00034   &  
        0.0086    &  
        0.0043    &  
        0.0024    &  
        0.0024    &  
        0.026     &  
        0.033     &  
        5.4       &  
        0.024        \\
\hspace{5pt} + $C_\l^{\psi g}$ 
 &
        0.011     &  
        0.00034   &  
        0.0086    &  
        0.0043    &  
        0.0024    &  
        0.0024    &  
        0.026     &  
        0.033     &  
        5.4       &  
        0.024        \\
\hspace{5pt} + $C_\l^{g\gamma}$ 
 &
        0.011     &  
        0.00031   &  
        0.0059    &  
        0.0043    &  
        0.0024    &  
        0.0023    &  
        0.026     &  
        0.022     &  
        5.2       &  
        0.024        \\
\hspace{5pt} + $C_\l^{Tg}$ + $C_\l^{\psi \gamma}$ 
 + $C_\l^{\psi g}$ + $C_\l^{g\gamma}$ 
 &
        0.011     &  
        0.00031   &  
        0.0059    &  
        0.0043    &  
        0.0024    &  
        0.0023    &  
        0.026     &  
        0.022     &  
        5.1       &  
        0.024        \\
\hline \hline
{\bf HSC + Planck + ACTPol} \\
CMB + $C_\l^{\rm gg}$ + $C_\l^{\psi \psi}$ + $C_\l^{\gamma \gamma}$ 
 &
        0.0074    &  
        0.00033   &  
        0.0086    &  
        0.0039    &  
        0.0022    &  
        0.0022    &  
        0.024     &  
        0.032     &  
        5.4       &  
        0.021        \\
\hspace{5pt} + $C_\l^{Tg}$  
 &
        0.0074    &  
        0.00033   &  
        0.0086    &  
        0.0039    &  
        0.0022    &  
        0.0022    &  
        0.024     &  
        0.032     &  
        5.3       &  
        0.021        \\
\hspace{5pt} + $C_\l^{\psi \gamma}$ 
 &
        0.0073    &  
        0.00032   &  
        0.0085    &  
        0.0039    &  
        0.0022    &  
        0.0022    &  
        0.024     &  
        0.032     &  
        5.4       &  
        0.021        \\ 
\hspace{5pt} + $C_\l^{\psi g}$ 
 &
        0.0073    &  
        0.00033   &  
        0.0085    &  
        0.0039    &  
        0.0022    &  
        0.0022    &  
        0.024     &  
        0.032     &  
        4.8       &  
        0.021        \\
\hspace{5pt} + $C_\l^{g\gamma}$ 
 &
        0.0073    &  
        0.00030   &  
        0.0059    &  
        0.0039    &  
        0.0021    &  
        0.0022    &  
        0.024     &  
        0.022     &  
        5.1       &  
        0.021        \\
\hspace{5pt} + $C_\l^{Tg}$ + $C_\l^{\psi \gamma}$ 
 + $C_\l^{\psi g}$ + $C_\l^{g\gamma}$ 
 & 
        0.0073    &  
        0.00029   &  
        0.0058    &  
        0.0039    &  
        0.0021    &  
        0.0022    &  
        0.024     &  
        0.022     &  
        4.8       &  
        0.021        
\end{tabular}
\end{ruledtabular}
\caption{Expected marginalized errors (1$\sigma$) and these forecasts
  correspond to the results of Fig.~\ref{fig:cont_cross}.}
\label{tb:cross}
\end{table}


\begin{table}[t]
\begin{ruledtabular}
\begin{tabular}{l||ccccc|ccccc}
\multicolumn{1}{c}{} & 
\multicolumn{5}{c}{HSC + Planck} & 
\multicolumn{5}{c}{HSC + Planck + ACTPol} \\
\hline \hline 
 &
$\Omega_\Lambda$ & $f_\nu$ & $w$ & $f_{\rm NL}$ & $\log M_{\rm obs}$ &
$\Omega_\Lambda$ & $f_\nu$ & $w$ & $f_{\rm NL}$ & $\log M_{\rm obs}$ \\
\hline
CMB + $C_\l^{\rm gg}$ + $C_\l^{\psi \psi}$ + $C_\l^{\gamma \gamma}$ &
0.015  &  0.0028  &  0.054  &  5.5  &  0.033 &
0.015  &  0.0023  &  0.053  &  5.4  &  0.029   
\\
\hspace{5pt} + $C_\l^{Tg}$  &
0.015  &  0.0028  &  0.053  &  5.4  &  0.033 &
0.014  &  0.0023  &  0.053  &  5.3  &  0.029   
\\
\hspace{5pt} + $C_\l^{\psi \gamma}$ &
0.015  &  0.0028  &  0.053  &  5.5  &  0.033 &
0.014  &  0.0023  &  0.053  &  5.4  &  0.029   
\\
\hspace{5pt} + $C_\l^{\psi g}$ &
0.015  &  0.0028  &  0.053  &  5.4  &  0.033 &
0.014  &  0.0023  &  0.051  &  4.8  &  0.028   
\\
\hspace{5pt} + $C_\l^{g\gamma}$ &
0.0095  &  0.0028  &  0.035  &  5.2  &  0.033 &
0.0095  &  0.0023  &  0.034  &  5.2  &  0.029   
\\
All &
0.0095  &  0.0028  &  0.034  &  5.1  &  0.033 &
0.0092  &  0.0023  &  0.033  &  4.8  &  0.028   
\end{tabular}
\end{ruledtabular}
\caption{ Expected marginalized errors (1$\sigma$) which correspond to the
  results of Fig.~\ref{fig:cont_sys} with all systematics.  This table should
  be compared with TABLE \ref{tb:cross} where the systematics are neglected.  }
\label{tb:sys}
\end{table}

\end{widetext}


\subsection{Contribution of cross correlations}

Now let us discuss the contribution from ACTPol experiment which will
extract the information of CMB lensing more efficiently than Planck.

As shown Fig.~\ref{fig:cont_lens} and Table~\ref{tb:lens}, we find that almost
all the constraints on each plane are determined through galaxy lensing
information except for $M_{\rm obs}$ and $f_{\rm NL}$.  However, CMB lensing
also gives important contributions if the ACTPol experiment is included.
Focusing on the constraints on the $f_{\rm NL}$-$M_{\rm obs}$ plane, both the CMB
lensing and galaxy lensing signals improve the constrains slightly by
including their cross correlations, though the most of the constraints are
determined by the galaxy auto correlation.

We show how the cross correlations improve the constraints in
Fig.~\ref{fig:cont_cross} and Table~\ref{tb:cross}.  We find that most of the
constraints are mainly determined through the auto correlations on each plane
but we can still gain a little benefit from the cross correlations. By
including all cross correlations among CMB, galaxy distribution and galaxy
lensing shear, we can expect $5.6\%$ improvement for the constraint on $f_{\rm
  NL}$ with HSC and Planck, and $11.1\%$ improvement with HSC, Planck and
ACTPol.  In particular, the impact of the cross correlation between galaxy and
galaxy lensing (g$\gamma$) is significant, especially in the
$w$-$\Omega_\Lambda$ plane. This result is consistent with that obtained by
\cite{Namikawa11,Oguri11} and g$\gamma$ will play a very important role in the
constraint on dark energy parameters.


Finally, to see the impact of cross correlations for the constraint on the
primordial non-Gaussianity, let us focus on the constraints on the $f_{\rm
  NL}$-$M_{\rm obs}$ plane in Fig.~\ref{fig:cont_cross}.  We find that the
cross correlation between galaxy and CMB lensing (g$\psi$) will improve the
constraint on $f_{\rm NL}$ when the Planck and ACTPol experiments are
combined. However we have little benefit from the cross correlations if we use
only Planck for the CMB experiment.


\subsection{Dependence on mass threshold $M_{\rm obs}$}

\begin{figure}[t]
  \includegraphics[clip,keepaspectratio=true,width=0.48
     \textwidth]{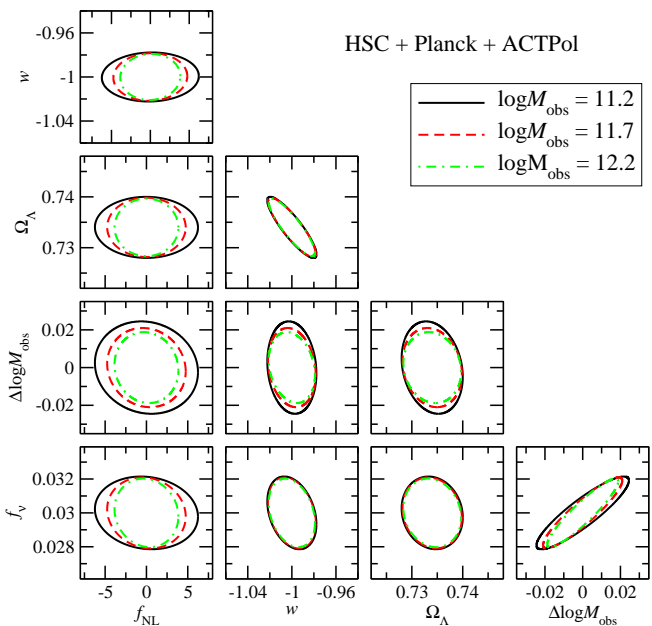}
\caption{Projected 1$\sigma$(68\%) confidence constraints in some parameter
  spaces for three fiducial values of the mass threshold $M_{\rm obs}
  [h^{-1}M_{\odot}] =10^{11.2},10^{11.7},10^{12.2}$.  We show the contours
  expected from combining HSC, Planck and ACTPol, and these results include
  all the information from the auto and cross correlations. Only in this
  figure, we plot the deviation from the fiducial values $\Delta \log M_{\rm
    obs}$ instead of $\log M_{\rm obs}$ for the purpose of illustrating the
  different fiducial models on the same planes. }
\label{fig:cont_Mobs}
\end{figure}

The parameter $M_{\rm obs}$ reflects the mass threshold we can observe as
galaxy and relates to the galaxy bias through Eq.~(\ref{eq:beff}). We can
interpret from Fig.~\ref{fig:beff} that the larger value of $M_{\rm obs}$
predicts the larger bias. We show the constraints for the different fiducial
value of $M_{\rm obs}$ in Fig.~\ref{fig:cont_Mobs}.  Only in this figure, we
plot $\Delta \log M_{\rm obs}$ instead of $\log M_{\rm obs}$ for the purpose
of illustrating the different fiducial models on the same planes.

The difference which comes from the different fiducial values of $M_{\rm obs}$
appears especially in the constraints on $f_{\rm NL}$ and $M_{\rm obs}$, while
the constraints on the other parameters are not altered so much. This is
because the constraints on $f_{\rm NL}$ and $M_{\rm obs}$ come mainly from the
galaxy distribution (g), while the constraints of the others come mainly from
the other observables. As for the constraints on $f_{\rm NL}$, the larger
value of $M_{\rm obs}$ can lead to a tighter constraint on $f_{\rm NL}$. The
larger value of $M_{\rm obs}$ means picking out the higher mass
objects. Because high mass objects exhibit a large bias, the non-Gaussian
correction to the bias becomes large. Furthermore, the correction to the mass
function also becomes large for high mass objects. For these reasons, the
effect of the non-Gaussianity through the effective bias $b_{\rm eff}$ becomes
clearer when we choose the larger values of $M_{\rm obs}$ for the fiducial
model. Thus we can constrain $f_{\rm NL}$ more tightly for larger the value of
$M_{\rm obs}$.


\begin{figure}[t]
  \includegraphics[clip,keepaspectratio=true,width=0.48
     \textwidth]{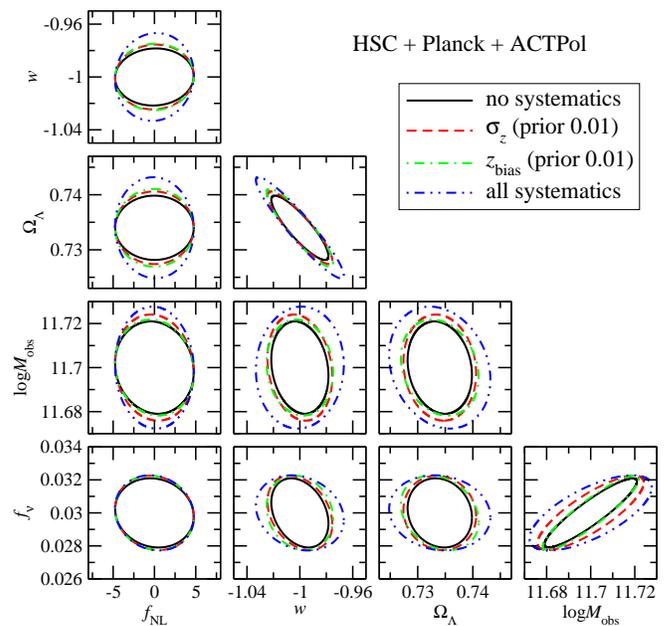}
\caption{Projected 1$\sigma$(68\%) confidence constraints in some parameter
  spaces without photometric redshift systematics, including the effects of
  scatter and bias individually, and including the both of the effects
  simultaneously.  The contours are obtained from HSC, Planck and ACTPol, and
  the results include the information from all auto and cross correlations.}
\label{fig:cont_sys}
\end{figure}



\subsection{Effects of photometric redshift systematics}

Here, we investigate effects of photometric redshift systematics on the
parameter constraints following examples in \cite{Das11}.

In our analysis, we assume the fiducial redshift scatter and bias as
Eqs.~(\ref{eq:sys_scatter}) and (\ref{eq:sys_bias}), respectively.  To
characterize the uncertainty in the scatter and bias, we parameterize
$\sigma_{z}^{(i)}$ and $z_{\rm bias}^{(i)}$ ($i=\{1,2,3,4,5\}$) in 5 bins,
include these 10 parameters in the Fisher matrix, apply a prior of $0.01$ on
each, and finally marginalize over these parameters.  We show the effects of
the photometric redshift systematics on constraints on some parameters in
Fig.~\ref{fig:cont_sys} and Table~\ref{tb:sys}.

In Fig.~\ref{fig:cont_sys}, we show the total effects of the scatter and bias
on the parameter determination, as well as the individual effects.  The
scatter gives a stronger impact than the bias, and including both of the
two systematics dramatically enlarges the constraint contours, except for
$f_\nu$ and $f_{\rm NL}$. Perhaps, for $f_\nu$, the constraint comes mainly
from CMB and CMB lensing, so the effects of systematics in the LSS observables
do not contribute to the constraint.  For the constraint on $f_{\rm NL}$, the
effects of systematics have little degeneracies with the effect of $f_{\rm
  NL}$ because we can distinguish these effects through the observation of the
power spectra at large-scale region (low-$\l$ region).

In Table~\ref{tb:sys}, we show $1\sigma$ errors for some parameters in the
case with HSC and Planck experiments, with or without ACTPol, sorted by the
cross correlation spectrum used in the analysis.  The relative amount of
contributions from each cross correlation to the constraints is similar to the
case without systematics.  However, we enjoy the benefits of including the
extra cross correlations more in the case with systematics than without.  For
example, information from the cross correlation brings $38.7\%$ and $37.7\%$
improvements of the constraints on $\Omega_\Lambda$ and $w$ in the case with
systematics, respectively, while bringing $32.6\%$ and $31.3\%$ improvements in the
case without systematics.


\begin{figure}[t]
  \includegraphics[clip,keepaspectratio=true,width=0.48
     \textwidth]{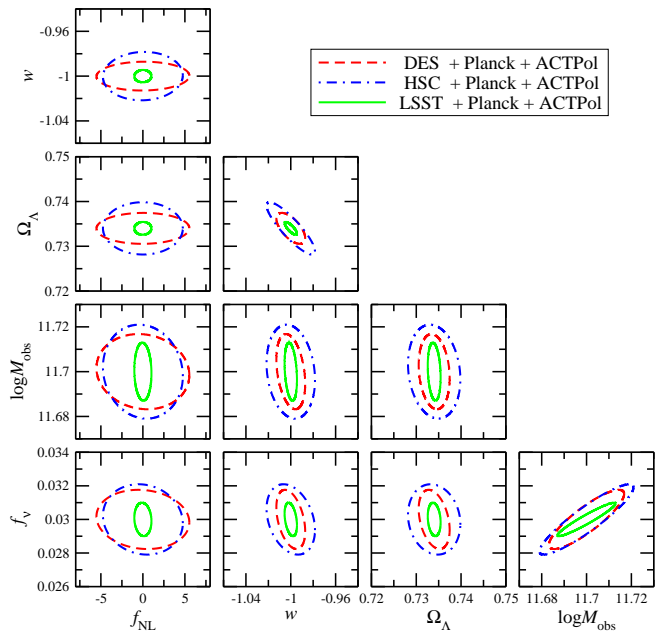}
\caption{Projected 1$\sigma$(68\%) confidence constraints in some parameter
  spaces for $f_{\nu}\neq 0$ model with the photometric redshift scatter. The
  contours are for DES (red dashed), HSC (blue dot-dashed) or LSST (green),
  with Planck and ACTPol combined. The results include all the information
  from auto and cross correlations.}
\label{fig:cont_survey}
\end{figure}



\subsection{Comparison with other various future surveys}

Focusing on the constraints on the primordial non-Gaussianity from the galaxy
power spectrum or two-point correlation function using the scale-dependent
bias, the result significantly depends on the survey strategies. The signature
of the primordial non-Gaussianity grows with redshift and appears prominently
on the large-scale, so surveys observing high-redshift and wide area are
predictably effective for this purpose.  Here, we show the constraints
expected from other survey projects besides HSC, such as DES \cite{DES} and LSST
\cite{LSST} in Fig.~\ref{fig:cont_survey}.

Comparing the HSC survey with DES, the DES survey is more suitable for the
constraints on the parameters related to dark energy, such as $\Omega_\Lambda$
and $w$, than HSC. For the constraint on the primordial non-Gaussianity,
however, HSC is more suitable than DES. This is because HSC will observe
galaxies in higher redshift than DES and it allows us to follow the redshift
evolution of the scale-dependent bias due to the primordial non-Gaussianity,
even though the survey area of DES is larger than HSC.  On the other hand, the
constraints from LSST are tremendous for most of the parameter determinations.


\section{Summary} \label{sec:summary}

In this work, we estimated errors in the determination of cosmological
parameters for some future surveys with the Fisher matrix method, newly
including galaxy-galaxy lensing (g$\gamma$) and CMB lensing-galaxy lensing
($\psi \gamma$) cross correlations. In general, the extra information from
cross correlations allows us to estimate the cosmological parameters more
precisely.

As for the constraint on the $f_{\rm NL}$-$\log M_{\rm obs}$ plane,
galaxy-galaxy lensing (g$\gamma$) cross correlation improves the constraint
greatly.  On the other hand, galaxy-CMB lensing (g$\psi$) cross correlation
can also improve the constraint if the ACTPol experiment is included.

Although galaxy-galaxy lensing (g$\gamma$) cross correlation has a great
contribution for improving the constraint on $f_{\rm NL}$, we have to pay
attention to the systematics of galaxy lensing observation.  The photo-$z$
error diffuses the observed galaxy distribution of each redshift bin, and
changes the overall amplitude of the power spectrum. This behavior is similar
to the change of the parameter $\log M_{\rm obs}$. On the other hand, positive
(negative) $f_{\rm NL}$ enhances (decreases) the amplitude only on
large-scales. Therefore comparing these two parameters, the constraint on
$\log M_{\rm obs}$ is more affected by photo-$z$ error than that of $f_{\rm
  NL}$.

Here, we accounted only for the photo-$z$ error for systematics, but it is known
that there are other systematics for the galaxy lensing observables.  One of
them is a complication in the measurement of the shear due to the incomplete
removal of the effects of the point spread function (PSF) \cite{Huterer06}.
So the contribution of galaxy-galaxy lensing (g$\gamma$) cross correlation to
the constraint on $f_{\rm NL}$ may be diminished if we consider a more
realistic condition including other systematics for galaxy lensing
observables. In such a case, the impact of galaxy-CMB lensing (g$\psi$) cross
correlation probably shines out more.

In general, the choice of the fiducial cosmological model affects the
parameter forecasts.  In this paper, we considered the differences of the
fiducial model in the neutrino mass and the photometric redshift scatter.  As
far as the constraint on $f_{\rm NL}$ is concerned, we found that the selection
of the fiducial model makes an impact on the result.

It should be emphasized that combining the satellite CMB experiment (Planck)
with the ground based one (ACTPol) can greatly improve the constraints on the
cosmological parameters.  This is because the information of CMB lensing can
be extracted more efficiently by combining Planck with ACTPol than by Planck
only. For the constraint on $f_{\rm NL}$, the marginalized error can be
improved from $\Delta f_{\rm NL} \sim 5.1$ to 4.8 by combining ACTPol with
Planck, because in this case galaxy-CMB lensing cross correlation (g$\psi$)
starts to play an important role in the improvement of the statistical error
of $f_{\rm NL}$.

As for the strategy of the survey, we found that HSC is preferable to DES for
the constraint on $f_{\rm NL}$ because the former can probe higher redshift.
We expect $\Delta f_{\rm NL} \sim 5.5$ with the DES survey while $\Delta
f_{\rm NL} \sim 4.8$ with HSC.  However, we never forget the importance of
wide field surveys to see the signature of the primordial non-Gaussianity
appearing on large-scale.

As for the benefits from the cross correlations, we found that the cross
correlation between CMB lensing and galaxy distribution improves the
constraints on $f_{\rm NL}$ from 5.4 to 4.8 $(\Delta 11.1\%)$, and from 1.1 to
1.0 $(\Delta 8.3\%)$ in the case with LSST.  The relative improvement of
$f_{\rm NL}$ by including cross correlations is less distinct in LSST, but the
cross correlation will be still significant in the future surveys.


\acknowledgments 

We thank O. Dor\'{e}, S. Dodelson, A. Lewis, T. Namikawa, and S. Yokoyama for useful
discussion and comments. We acknowledge support from JSPS (Japan Society for
Promotion of Science) Grant-in-Aid for Scientific Research under Grant
No. 22012004 (KI); Grant-in-Aid for Nagoya University Global COE Program
``Quest for Fundamental Principles in the Universe: from Particles to the
Solar System and the Cosmos'', from the Ministry of Education, Cluster,
Sports, Science, and Technology (MEXT); Kobayashi-Maskawa Institute for the
Origin of Particles and the Universe; Nagoya University for providing
computing resources useful in conducting the research reported in this paper,
Grant-in-Aid for Scientific Research (C), under Grant No. 21540263, 2009 (TM);
and Grant-in-Aid for Scientific Research on Priority Areas No. 467 ``Probing
the Dark Energy through an Extremely Wide and Deep Survey with Subaru
Telescope''. This work is supported in part by JSPS Core-to-Core Program
``International Research Network for Dark Energy''.

\bibliography{cmblss}

\end{document}